\documentclass[12pt]{article}
\newcommand{\myappendix}{\setcounter{equation}{0}\appendix}
\usepackage{epsfig}
\usepackage{graphicx}
\usepackage{a4}
\usepackage{latexsym}
\usepackage{cite}

\usepackage{pslatex}
\usepackage[latin1]{inputenc}
\usepackage[T1]{fontenc}
\newcommand{\ep}{\mbox{$\varepsilon$}}

\newcommand{\xo}{\mbox{$x_1^{\!{0}\!}$}}
\newcommand{\xt}{\mbox{$x_2^{\!{0}\!}$}}

\begin{document}
\setlength{\parskip}{0.2cm}
\setlength{\baselineskip}{0.55cm}
\begin{titlepage}
\begin{flushright} 
\hfill {\tt hep-ph/xxxxxxx}
\\
\hfill {\tt HRI-04/2006}
\\
\hfill {\tt YITP-SB-06-35}
\end{flushright} 
\vspace{5mm} 
\begin{center} 
{\Large \bf 
QCD threshold corrections to di-lepton and Higgs
rapidity distributions beyond N${}^2$LO  
}\\
\end{center}

\vspace{10pt} 
\begin{center} 
{\bf 
V.~ Ravindran$^1$, J.~ Smith${^2}$, W.L.~ van Neerven$^{3}$
}\\ 
\end{center}
\begin{center} 
{\it 
${}^1$Harish-Chandra Research Institute, 
 Chhatnag Road, Jhunsi, Allahabad, India,\\
${}^2$C.N. Yang Institute for Theoretical Physics,
Stony Brook University, Stony Brook, NY~11794-3840 USA,
\\${}^3$ Lorentz Institute, University of Leiden, P.O. Box 9502, 2300
 RA Leiden,The Netherlands.
} 
\end{center}
 
\vspace{10pt} 
\begin{center}
{\bf ABSTRACT} 
\end{center} 
We present threshold enhanced QCD corrections to rapidity 
distributions of di-leptons in the Drell-Yan process 
and of Higgs particles in both gluon fusion and
bottom quark annihilation processes using Sudakov 
resummed cross sections.  We have used
renormalisation group invariance and the mass factorisation theorem 
that these hard scattering cross sections satisfy  as well as 
Sudakov resummation of QCD amplitudes. 
We find that these higher order threshold QCD corrections
stabilise the theoretical predictions under scale variations. 
\vskip12pt 
\vskip 0.3 cm

\end{titlepage}
Perturbative Quantum Chromodynamics (pQCD) provides a framework  
to successfully compute various observables in the collisions of
hadrons at high energies.  Recent theoretical advances in the computations
of higher order QCD radiative corrections have lead to precise results for 
several important observables. Because of this progress, we can now predict
these observables with unprecedented accuracy for physics studies at the 
Tevatron collider in Fermilab as well as at the upcoming Large Hadron 
Collider (LHC) in CERN
\cite{Dittmar:2005ed}.
 
The Drell-Yan (DY) production of di-leptons \cite{Drell:1970wh}
has been one of the most important probes of the structure of hadrons.
It is also one of the dominant production processes at hadron colliders. 
At the LHC, it will serve as a luminosity monitor
which is very important to precisely calibrate the machine for 
searches for physics beyond the Standard Model (SM). 
In DY production, a pair of leptons is produced through the decay of
virtual photons, Z and W bosons that result from the collisions 
of incoming partons (quarks and gluons) in the hadrons. 
At hadron colliders, the DY process provides precise measurements of various
standard model parameters. Rapidity distributions of Z bosons
\cite{Affolder:2000rx}
and charge asymmetries of leptons coming from W boson decays
\cite{Abe:1998rv}
can probe the structure of the hadrons 
and possible excess events in di-lepton invariant mass distributions 
can point to physics beyond the standard model such as $R$-parity violating
supersymmetric models and models with Z${}'$, or with  
contact interactions \cite{Affolder:2001ha}.
Both D0 and CDF collaborations \cite{Patwa:2006rd}
at the Fermilab Tevatron made precise measurements of $Z$ and $W$ production 
cross sections and asymmetries which not only allowed for stringent tests 
of the standard model but also play an important
role in the Higgs search at future colliders.
These measurements are also possible at the LHC due to the large cross 
sections for the DY process.

The other process which is equally important is Higgs boson production 
at these colliders because it will establish the Standard Model as well as 
look for beyond the SM Higgs \cite{Djouadi:2005gi,Djouadi:2005gj}.
The Higgs boson, which is responsible for the electroweak symmetry breaking in
the Standard Model, is yet to be discovered.  The search for this particle
has been going on at the Fermilab Tevatron and is one of the most important
tasks for the CERN LHC.  The LEP experiments in the past provided vital 
information on the possible mass range of this particle \cite{Barate:2003sz}.   
The lower bound on the mass is $114.4$ GeV$/c^2$ and an upper bound 
is less than $219$ GeV$/c^2$ at $95\%$ CL \cite{:2004qh}.  At the LHC, 
Higgs bosons will be predominantly produced through the gluon fusion process 
due to the large flux of gluons in the protons at these energies.  They can be
detected through the rare two photon decay mode which has less QCD
background than other signals.  

In pQCD, the total cross sections for the DY production of di-leptons and 
Higgs boson production are known upto next-to-next-to-leading order (NNLO) 
level
\cite{Kubar-Andre:1978uy,Altarelli:1978id,Humpert:1980uv,
Dawson:1990zj,Djouadi:1991tk,Spira:1995rr,Matsuura:1987wt,
Matsuura:1988sm,Hamberg:1990np,Harlander:2001is,Catani:2001ic,Catani:2003zt, 
Harlander:2002wh,Anastasiou:2002yz,Ravindran:2003um,Harlander:2003ai,
Ravindran:2004mb}.
However, due to the complexity involved with the top quark loops, 
the Higgs production cross sections are only known 
in the large top quark mass limit beyond the next-to-leading order (NLO). 
In addition to these fixed order results, the resummation programs 
for the threshold corrections 
to the total cross sections for DY and Higgs production have 
also been very successful
\cite{Sterman:1986aj,Catani:1989ne}(see also \cite{Kodaira:1981nh}).
See \cite{Vogt:2000ci,Catani:2003zt}
for next-to-next-to-leading logarithmic (NNLL) resummation results. 
Due to  several important QCD results at the three loop level 
that are recently available
\cite{Moch:2004pa,Vogt:2004mw,Moch:2005id,Moch:2005tm,Vermaseren:2005qc,
Moch:2005ba,
Blumlein:2004xt},
the resummation upto N${}^3$LL has also become a reality \cite{Moch:2005ky,
Laenen:2005uz,Idilbi:2005ni,Ravindran:2005vv}. 
The fixed order partial soft-plus-virtual N${}^3$LO corrections
\cite{Moch:2005ky,Ravindran:2006cg} 
to the DY and Higgs productions show the reliability of the perturbation
theory and the stability against the scale variations.
The fixed order results as well as the resummed results reveal very  
interesting structures in the perturbative QCD results (see, 
\cite{Blumlein:2000wh,Blumlein:2006pj,
Blumlein:2005im,Dokshitzer:2005bf,Ravindran:2005vv}).

Infra-red safe observables, such as hadronic cross sections, are computed
using the QCD improved parton model. 
Due to the factorisation property that certain hard scattering cross sections
satisfy, they can be expressed interms of finite partonic cross sections
convoluted with the parton distributions functions (PDFs).  The partonic 
cross sections are calculated in QCD using standard perturbation theory
in powers of the strong coupling constant $g_s$ that becomes small at 
high energies.  The ultraviolet singularities that arise beyond 
leading order are often removed in the ${\overline {\rm MS}}$ renormalisation 
scheme at a renormalisation scale $\mu_R$. The collinear singularities 
that result due to the presence of light mass partons are mass factorised 
into the bare parton densities in the ${\overline {\rm MS}}$ scheme at a 
factorisation scale $\mu_F$.  Hence the fixed order perturbative results 
are often sensitive to these scales $\mu_R$ and $\mu_F$.  For example
see \cite{Cafarella:2005nb} for a study of scale variations of the 
Higgs cross section in NLO. However results that are 
known to sufficiently high order in the strong coupling constant are 
often less sensitive to these scales because the observables are 
renormalisation group invariant. That is they would be strictly independent 
of choice of these scales if the entire perturbative expansion were known.

In addition to the scale uncertainties the fixed order computations
suffer from the presence of various large logarithms which arise in some 
kinematical regions.  These regions are often important from the experimental
point of view.  The large largarithms spoil the standard perturbative approach.
The alternate approach is to resumm these logarithms in a closed form.
Resumming a class of large logarithms supplemented with 
fixed order results can usually cover the entire kinematic region of
the phase space.  In this paper, we will mainly concentrate on a class of 
logarithms that arise in the threshold regions.  These threshold corrections 
are further enhanced when the fluxes of the incoming partons become large 
in those regions.  In the case of Higgs production through gluon fusion, 
the gluon flux at small partonic energies becomes large enhancing the role of
threshold corrections.  Here we examine the effects of soft gluons that 
originate in the threshold region of the phase space when we consider 
the $x_F$ and rapidity distributions in DY production and 
Higgs production through both gluon fusion and bottom quark annihilation. 
Here the large logarithms are generated when the gluons
that are emitted from the incoming/outgoing partons become soft.   

In \cite{Ravindran:2005vv}, we found that 
soft distribution functions of Drell-Yan and Higgs production
cross sections in perturbative QCD are maximally non-abelian.
That is, we found that the soft distribution function for
Higgs production can be obtained entirely
from the DY process by a simple multiplication of the colour
factor $C_A/C_F$.   In the article \cite{Ravindran:2006cg}, 
using the soft distribution functions extracted
from DY, and the form factor of the Yukawa coupling of Higgs to
bottom quarks, we predicted the soft-plus-virtual (sv) part of
the Higgs production through bottom quark annihilation beyond NNLO
with the same accuracy that the DY process and the gluon fusion to
Higgs process are known \cite{Moch:2005ky,Ravindran:2005vv}.
We extended \cite{Blumlein:2006pj} this approach to entirely different 
processes such as Higgs decay to bottom quarks and hadroproduction 
in $e^+e^-$ annihilation.  The approach that we followed 
in \cite{Ravindran:2005vv,Ravindran:2006cg,Blumlein:2006pj} is closely 
related to that of the standard threshold resummation and hence we could 
determine \cite{Ravindran:2005vv} the threshold exponents $D_i^I$ upto 
three loop level for DY and Higgs production using our resummed soft 
distribution functions and $B_i^I$ for both deep inelastic scattering
and Higgs decay and hadroproduction.  In this paper we extend this approach 
to include differential cross sections such as $x_F$ and rapidity 
distributions of the di-lepton pair in DY production
and of Higgs bosons in Higg production processes.

In the following we systematically formulate a framework to resum the 
dominant soft gluon contributions to these differential cross sections.
We perform the resummation in the $z_i(i=1,2)$ space of the kinematic 
variables, which are the appropriate scaling variables that enter the 
differential partonic cross sections.  
The threshold region corresponds to $z_i \rightarrow 1$ and in this region
all the partonic cross sections are symmetric in $z_1 \leftrightarrow z_2$.
We have used renormalisation group (RG) invariance, mass factorisation and 
Sudakov resummation of QCD amplitudes as guiding principles to perform the 
resummation in this region.  Using the resummed results in $z_i$ space 
we predict the soft-plus-virtual parts (also called threshold corrections) 
of the dominant partonic differential cross sections beyond N${}^2$LO.  
We also study the numerical effect of our predictions on both the $x_F$ and 
rapidity distributions of di-leptons and Higgs bosons.  
The analytical results are presented in the Appendices for both DY and Higgs 
production 
\footnote{The results for the Higgs 
production via bottom quark annihilation are not presented here but can be
obtained from the authors on request.}
through gluon fusion. For an early reference where the resummation for DY 
differential distributions at rapidity $Y=0$ (or $x_F=0$) was 
considered consult 
\cite{laenensterman}.

The differential cross section can be expressed as:
\begin{eqnarray}
{d \sigma^I \over dx } = 
\sigma^I_{\rm Born}(x_1^0,x_2^0,q^2) W^I(x_1^0,x_2^0,q^2) ~,
\quad \quad \quad I=q,b,g \quad\,,
\end{eqnarray}
with the normalisation 
$W^I_{\rm Born}(x_1^0,x_2^0,q^2)=\delta(1-x_1^0)\delta(1-x_2^0)$.
The $x^0_{i}~(i=1,2)$ are related to the kinematical variables $q^2$ and $x$.  
Here $q$ is the momentum of the di-lepton pair in the DY process
and of the Higgs boson in the Higgs production.
The variable $x$ can be the $x_F$ or rapidity of the di-lepton pair 
or of the Higgs boson.
For di-lepton production, $I=q$ and $\sigma^I=d\sigma^{q}(\tau,q^2,x)/dq^2$ 
with $q^2$ the invariant mass of the final state di-lepton pair 
i.e, $q^2=M^2_{l^+l^-}$.  For Higgs production through gluon 
fusion, $I=g$ and $\sigma^I=\sigma^g(\tau,q^2,x)$ and 
for Higgs production through bottom quark annihilation
$I=b$ and $\sigma^I=\sigma^b(\tau,q^2,x)$ with $q^2=m_H^2$ where $m_H$ is 
the mass of the Higgs boson.  The variable $\tau=q^2/S$ with  $S=(p_1+p_2)^2$
the center of mass energy squared where $p_i$ are the momenta of 
incoming hadrons $P_i~(i=1,2)$.
In the QCD improved parton model, the function $W^I(x_1^0,x_2^0,q^2)$ 
can be expressed in terms of the PDFs appropriately
convoluted with perturbatively calculable partonic differential cross sections
denoted by $\Delta^I_{d,ab}$, where the subscript $"d"$ stands for 
"differential", as follows
\begin{eqnarray}
W^I(x_1^0,x_2^0,q^2) &=&\sum_{ab=q,\overline q,g} 
\int_0^1 dx_1 \int_0^1 dx_2~ {\cal H}^I_{ab}(x_1,x_2,\mu_F^2) 
\nonumber\\[2ex]
&&\times \int_0^1 dz_1 \int_0^1 dz_2
~\delta(x_1^0-x_1 z_1)~ \delta(x_2^0-x_2 z_2)~ 
\Delta^I_{d,ab} (z_1,z_2,q^2,\mu_F^2,\mu_R^2) \,.
\end{eqnarray}
Here, $\mu_R$ is the renormalisation scale and $\mu_F$ the factorisation scale.
We consider the differential cross sections for two kinematic variables namely
\begin{eqnarray}
x=x_F={2 (p_1-p_2)\cdot q \over S}, \quad \quad &{\rm and}& \quad 
\quad  x=Y={1 \over 2} 
\log\left({p_2 \cdot q \over p_1 \cdot q}\right) \,.
\end{eqnarray}
For the $x_F$ ($x=x_F$) distribution, the $x_i^0$ variables satisfy
\begin{eqnarray}
x_F=x_1^0-x_2^0, \quad \quad \quad \tau=x_1^0 x_2^0~\,,
\end{eqnarray}
while for the rapidity $Y$ ($x=Y$) distribution,  we have
\begin{eqnarray}
Y={1 \over 2 } \log\left({x_1^0 \over x_2^0}\right), \quad \quad \quad \tau=x_1^0 x_2^0 ~.
\end{eqnarray}
Here, the function ${\cal H}^I_{ab}(x_1,x_2,\mu_F^2)$ is the product of PDFs
$f_a(x_1,\mu_F^2)$ and $f_b(x_2,\mu_F^2)$ renormalised at the
factorisation scale $\mu_F$.  That is,
\begin{eqnarray}
{\cal H}^q_{ab}(x_1,x_2,\mu_F^2)&=& 
f^{P_1}_a(x_1,\mu_F^2)~ f^{P_2}_b(x_2,\mu_F^2)\,,
\nonumber\\[2ex]
{\cal H}^g_{ab}(x_1,x_2,\mu_F^2)&=& 
x_1~ f^{P_1}_a(x_1,\mu_F^2)~ x_2~ f^{P_2}_b(x_2,\mu_F^2)\,,
\end{eqnarray}
with $x_i~(i=1,2)$ the momentum fractions of the partons in the 
incoming hadrons.

We first study the contributions coming from the soft gluons. 
The infra-red safe contributions from the soft gluons can be obtained by adding the soft part of
the differential cross sections with the ultraviolet renormalised virtual contributions and performing 
mass factorisation using appropriate counter terms.  
This combination is called the "soft-plus-virtual" (sv) 
part of the differential cross section. We call the remaining part the 
hard part of the differential cross section.  Hence we write
\begin{eqnarray}
\Delta^I_{d,ab} (z_1,z_2,q^2,\mu_F^2,\mu_R^2) = 
\Delta^{{\rm hard}}_{I,ab}(z_1,z_2,q^2,\mu_F^2,\mu_R^2)
+\delta_{a\overline b} \Delta^{\rm sv}_{~d,I}(z_1,z_2,q^2,\mu_F^2,\mu_R^2), 
\quad \quad \quad I=q,b,g \,.
\end{eqnarray}
The contributions coming from the hard parts 
$\Delta^{\rm hard}_{I,ab}(z_1,z_2,q^2,\mu_F^2,\mu_R^2)$ of the differential 
cross sections can be obtained by the standard procedure discussed in detail in 
\cite{Rijken:1994sh,Mathews:2004xp}.
The soft-plus-virtual parts of the differential cross sections
($\Delta^{\rm sv}_{~d,I}(z_1,z_2,q^2,\mu_R^2,\mu_F^2)$)
are found to be
\begin{eqnarray}
\Delta^{\rm sv}_{~d,I}(z_1,z_2,q^2,\mu_R^2,\mu_F^2)={\cal C} \exp
\Bigg({\Psi^I_d(q^2,\mu_R^2,\mu_F^2,z_1,z_2,\ep)}\Bigg)\Bigg|_{\ep=0}\,,
\label{master}
\end{eqnarray}
where $\Psi^I_d(q^2,\mu_R^2,\mu_F^2,z_1,z_2,\ep)$ are finite distributions
They are computed in $4+\ep$ dimensions and take the form
\begin{eqnarray}
\Psi^I_d(q^2,\mu_R^2,\mu_F^2,z_1,z_2,\ep)&=&
\Bigg(
\ln \Big(Z^I(\hat a_s,\mu_R^2,\mu^2,\ep)\Big)^2
+\ln \big|\hat F^I(\hat a_s,Q^2,\mu^2,\ep)\big|^2
\Bigg)
\delta(1-z_1) \delta(1-z_2)
\nonumber\\[2ex]
&&+2~ \Phi^{~I}_d(\hat a_s,q^2,\mu^2,z_1,z_2,\ep)
- {\cal C}\ln \Gamma_{II}(\hat a_s,\mu^2,\mu_F^2,z_1,\ep)~ \delta(1-z_2)
\nonumber\\[2ex]
&&- {\cal C}\ln \Gamma_{II}(\hat a_s,\mu^2,\mu_F^2,z_2,\ep)~ \delta(1-z_1)
\,,\hspace{2cm} I=q,b,g\,. 
\label{DYH}
\end{eqnarray}
The symbol "${\cal C}$" 
means convolution.  For example, ${\cal C}$ acting on the exponential
of a function $f(z_1,z_2)$ means the following expansion:
\begin{eqnarray}
{\cal C}e^{\displaystyle f(z_1,z_2) }&=& \delta(1-z_1)\delta(1-z_2)  + {1 \over 1!} f(z_1,z_2)
 +{1 \over 2!} f(z_1,z_2) \otimes f(z_1,z_2) 
\nonumber\\[2ex]
&& + {1 \over 3!} f(z_1,z_2) \otimes f(z_1,z_2) 
 \otimes f(z_1,z_2)
+ \cdot \cdot \cdot \,.
\end{eqnarray}
In the rest of the paper the function $f(z_1,z_2)$ 
is a distribution of the kind $\delta(1-z_j)$ or ${\cal D}_i(z_j)$, where
\begin{eqnarray}
{\cal D}_i(z_j)=\Bigg[{\ln^i(1-z_j) \over (1-z_j)}\Bigg]_+
\quad \quad \quad i=0,1,\cdot\cdot\cdot ,\quad {\rm and} \quad  j=1,2 \,,
\end{eqnarray}
and the symbol $\otimes$ means the "double" Mellin convolution. 
It convolutes  with respect to the variables $z_1$ and $z_2$ separately.
Since we are only interested in the sv part of the
cross sections, we drop all the regular functions that result from
various convolutions.
$\hat F^I(\hat a_s,Q^2,\mu^2,\ep)$ are the form factors
that contribute to di-lepton ($I=q$) (in DY) and Higgs ($I=g,b$) production. 
In the form factors, we have $Q^2=-q^2$.
The partonic cross sections depend on two scaling variables $z_1,z_2$. 
The functions $\Phi^{~I}_d(\hat a_s,q^2,\mu^2,z_1,z_2,\ep)$ are
called the soft distribution functions.  
The unrenormalised (bare) strong coupling constant
$\hat a_s$ is defined as
\begin{eqnarray}
\hat a_s={\hat g^2_s \over 16 \pi^2}\,,
\end{eqnarray}
where $\hat g_s$ is the strong coupling constant which is dimensionless in
$n=4+\ep$ space time dimensions.  The scale $\mu$ comes from the 
dimensional regularisation which makes the bare coupling constant $\hat g_s$
dimensionless in $n$ dimensions.
The bare coupling constant $\hat a_s$ is related to renormalised one by
the following relation:
\begin{eqnarray}
S_{\ep} \hat a_s = Z(\mu_R^2) a_s(\mu_R^2) 
\left(\mu^2 \over \mu_R^2\right)^{\ep \over 2}\,,
\label{renas}
\end{eqnarray}
where $S_{\ep}=\exp\left\{{\ep \over 2} [\gamma_E-\ln 4\pi]\right\}$
is the spherical factor characteristic of $n$-dimensional regularisation.
The renormalisation constant $Z(\mu_R^2)$ relates the bare coupling constant
$\hat a_s$ to the renormalised one $a_s(\mu_R^2)$.
They are both expressed in terms of the perturbatively calculable 
coefficients $\beta_i$ which are known up to four-loop level in terms of
the colour factors of SU(N) gauge group:
\begin{eqnarray}
C_A=N,\quad \quad \quad C_F={N^2-1 \over 2 N} , \quad \quad \quad
T_F={1 \over 2} \,.
\end{eqnarray}
Also we use $n_f$ for the number of active flavours.  In the case of Higgs
production, the number of active flavours is five because the
top degree of freedom is integrated out in the large $m_{\rm top}$ limit.

The factors $Z^I(\hat a_s,\mu_R^2,\mu^2,\ep)$ are the overall
operator renormalisation constants.
For the vector current  $Z^q(\hat a_s,\mu_R^2,\mu^2,\ep)=1$,
but both the gluon operator 
\cite{Chetyrkin:1997un} and the bottom quark coupling to Higgs
\cite{vanRitbergen:1997va}
get overall renormalisations.  They satisfy the following RG equations:
\begin{eqnarray}
\mu_R^2 {d \over d\mu_R^2} \ln Z^g(\hat a_s,\mu_R^2,\mu^2,\ep) &=&
\sum_{i=1}^\infty a^i_s(\mu_R^2)~ \Big(i~\beta_{i-1}\Big)\,,
\nonumber\\
\mu_R^2 {d \over d\mu_R^2}\ln Z^b(\hat a_s,\mu_R^2,\mu^2,\ep)&=&
\sum_{i=1}^\infty a^i_s(\mu_R^2)~ \gamma^b_{i-1}\,,
\label{ZRG}
\end{eqnarray}
where the limit $\ep \rightarrow 0$ is taken.  The constants
$i~\beta_{i-1}$ and $\gamma^b_{i-1}$ are the anomalous dimensions
of the renormalised form factors $F^g$ and $F^b$ respectively.

The bare form factors $\hat F^I(\hat a_s,Q^2,\mu^2,\ep)$
of both fermionic and gluonic operators satisfy
the following differential equation that follows from the gauge
as well as the renormalisation group 
invariances \cite{Sudakov:1954sw,Mueller:1979ih,
Collins:1980ih,Sen:1981sd}.  In dimensional regularisation,
\begin{eqnarray}
Q^2{d \over dQ^2} \ln \hat {F^I}\left(\hat a_s,Q^2,\mu^2,\ep\right)&=&
{1 \over 2 }
\Bigg[K^I\left(\hat a_s,{\mu_R^2 \over \mu^2},\ep\right)
+ G^I\left(\hat a_s,{Q^2 \over \mu_R^2},{\mu_R^2 \over \mu^2},\ep\right)
\Bigg]\,,
\label{sud1}
\end{eqnarray}
where the $K^I$ contain all the poles in $\ep$ and the
$G^I$ collect the rest of the terms that are
finite as $\ep$ becomes zero.  The fact that the 
$\hat F^I(\hat a_s,Q^2,\mu^2,\ep)$ are  renormalisation group 
invariant and the functions $G^I$ are finite implies that
the $K^I$ terms can be expressed in terms of 
finite constants $A^I$, the so-called cusp anomalous dimensions and the
coefficients $\beta_i$.
The solution to the eqn.(\ref{sud1}) can be obtained as a series expansion in
the bare coupling constant in dimensional regularisation.  The formal solution
up to four-loop level can be found 
in \cite{Moch:2005id,Ravindran:2005vv}.

The boundary conditions on the Sudakov differential equation, 
denoted by $G^{~I}_i(\ep)$ (see eqn.(19) of \cite{Ravindran:2006cg})
can be found for both $I=q$ and $I=g$ in \cite{Moch:2005tm} to the 
required accuracy in $\ep$.  We have extended this 
in \cite{Ravindran:2006cg} to the form factor 
corresponding to the Yukawa interaction of the Higgs boson to the 
bottom quarks.  
These constants  $G^{~I}_i(\ep)$  are expressed in terms of the functions 
$B_i^I$ and $f_i^I$. The $B_i^I$ are known up to order $a_s^3$  
through the three-loop anomalous dimensions (or splitting functions)
\cite {Moch:2004pa,Vogt:2004mw} and are found to be
flavour independent, that is $B_i^q=B_i^b$.
The constants $f_i^I$ are analogous to the cusp anomalous dimensions
$A_i^I$ that enter the form factors with $A_i^q=A_i^b$.  
It was first noticed in
\cite{Ravindran:2004mb}
that the single pole (in $\ep$) of the logarithm of the form factors
upto two-loop level ($a_s^2$) can be predicted 
\footnote{A similar analysis of the structure of single pole terms
of four-point amplitutes at the two-loop level can be found in 
\cite{Aybat:2006wq,Aybat:2006mz}.}
due the presence of constants $f_i^I$ because these
$f^{~I}_i$ are found to be maximally non-abelian obeying the relation
\begin{eqnarray}
f_i^q=f_i^b={C_F\over C_A} f_i^g\,,
\end{eqnarray}
similar to the $A_i^I$.    
This relation has been found to hold even at the three loop level  
\cite{Moch:2005tm}.
With this information we can now
predict all the poles of the form factors at every order in $\hat a_s$
from these constants $A^I$,$B^I$,$f^I$, their anomalous dimensions, and the 
finite parts of the lower order (in $\hat a_s$) contributions to the 
form factors.
Interestingly, the single pole terms in the form factors contain
the combinations
\cite{ Ravindran:2004mb,Moch:2005id,Ravindran:2006cg,Ravindran:2005vv}
$$
2 \Bigg(B^{~I}_i - \delta_{I,g}~ i~\beta_{i-1} -\delta_{I,b}
\gamma^b_{i-1}\Bigg) + f^{~I}_i\,,
$$ 
at order $\hat a_s^i$.
The terms $-2\delta_{I,g}~i~ \beta_{i-1} -2\delta_{I,b} \gamma^b_{i-1}$ 
come from 
the ultraviolet divergences that are present in the loop integrals.  
These pole terms go away when the form factors undergo overall operator
UV renormalisation through the renormalisation constants
$Z^{~I}$ which satisfy the RG equations given in eq.(\ref{ZRG}).

The collinear singularities that arise due to massless partons are
removed using the mass factorisation kernel $\Gamma(z_j,\mu_F^2,\ep)$ 
in the $\overline {\rm MS}$ scheme (see eqn.(\ref{DYH})).   
We have suppressed the dependence on
$\hat a_s$ and $\mu^2$ in $\Gamma$.
The factorisation kernel $\Gamma(z_j,\mu_F^2,\ep)$ satisfies the 
following renormalisation group equation:  
\begin{eqnarray}
\mu_F^2 {d \over d\mu_F^2}\Gamma(z_j,\mu_F^2,\ep)={1 \over 2}  P
                         \left(z_j,\mu_F^2\right)
                        \otimes \Gamma \left(z_j,\mu_F^2,\ep\right)\,,
\end{eqnarray}
where the $P(z_j,\mu_F^2)$ are the well-known DGLAP matrix-valued 
splitting functions which are known upto three-loop 
level \cite{Moch:2004pa,Vogt:2004mw}:
\begin{eqnarray}
P(z_j,\mu_F^2)=
\sum_{i=1}^{\infty}a_s^i(\mu_F^2) P^{(i-1)}(z_j)\,.
\end{eqnarray}
The diagonal terms in the splitting functions $P^{(i)}(z_j)$ have the 
following structure
\begin{eqnarray}
P^{(i)}_{II}(z_j) = 2\Bigg[ B^I_{i+1} \delta(1-z_j)
         + A^I_{i+1} {\cal D}_0(z_j)\Bigg] + P_{{\rm reg},II}^{(i)}(z_j)\,,
\end{eqnarray}
where $P_{{\rm reg},II}^{(i)}(z_j)$ are regular when the argument approaches 
the kinematic limit (here $z_j \rightarrow 1$).
The RG equations can be solved by expanding in powers of the
strong coupling constant.  For the 
soft-plus-virtual part of the differential cross sections, 
only the diagonal parts of the kernels contribute.  
We find the solutions contain only poles in $\ep$ in the 
$\overline{\rm MS}$ scheme:
\begin{eqnarray}
\Gamma(z_j,\mu_F^2,\ep)=\delta(1-z_j)+\sum_{i=1}^\infty \hat a_s^i
\left({\mu_F^2 \over \mu^2}\right)^{i {\ep \over 2}}S^i_{\ep}
\Gamma^{(i)}(z_j,\ep)\,.
\end{eqnarray}
The constants $\Gamma^{(i)}(z_j,\ep)$ expanded in negative powers of
$\ep$ up to four-loop level can be found in \cite{Ravindran:2005vv}.
The $\Gamma_{II}(\hat a_s,\mu^2,\mu_F^2,z_j,\ep)$ in eqn.(\ref{DYH})
is the diagonal element of $\Gamma(z_j,\mu_F^2,\ep)$.

The fact that $\Delta^{\rm sv}_{~I}$ are finite in the limit 
$\ep \rightarrow 0$ implies
that the soft distribution functions should have a pole structure in $\ep$ 
similar to that of $\hat F^I$ and $\Gamma_{II}$.  
To systematically study the soft distribution functions, we demand that 
they satisfy similar Sudakov type differential equations that
the form factors $\hat F^I$ satisfy (see eqn.(\ref{sud1})):
\begin{eqnarray}
q^2 {d \over dq^2}\Phi^{~I}_d(\hat a_s,q^2,\mu^2,z_1,z_2,\ep) =
{1 \over 2 }
\Bigg[\overline K^{~I}_d\left(\hat a_s,{\mu_R^2 \over \mu^2},z_1,z_2,\ep\right)
+ \overline G^{~I}_d\left(\hat a_s,{q^2 \over \mu_R^2},
{\mu_R^2 \over \mu^2},z_1,z_2,\ep\right)
\Bigg]\,,
\label{sud2}
\end{eqnarray}
where again the constants $\overline K^{~I}_d$ contain all the 
singular terms in $\ep$ and the $\overline G^{~I}_d$
are finite functions of $\ep$.  
Also the functions $\Phi^{~I}_d(\hat a_s,q^2,\mu^2,z_1,z_2,\ep)$ satisfy
the renormalisation group equations:
\begin{eqnarray}
\mu_R^2 {d \over d\mu_R^2}\Phi^{~I}_d(\hat a_s,q^2,\mu^2,z_1,z_2,\ep)=0\,.
\end{eqnarray}
This renormalisation group invariance leads to the following equations
\begin{eqnarray}
\mu_R^2 {d\over d\mu_R^2} \overline K^{~I}_d
\Bigg(\hat a_s, {\mu_R^2 \over \mu^2},z_1,z_2,\ep\Bigg)=
-\overline A^{~I}(a_s(\mu_R^2)) \delta(1-z_1)\delta(1-z_2) \,,
\nonumber\\[2ex]
\mu_R^2 {d \over d\mu_R^2} \overline G^{~I}_d
\Bigg(\hat a_s,{q^2 \over \mu_R^2},{\mu_R^2 \over \mu^2},z_1,z_2,\ep\Bigg)
=\overline A^{~I}(a_s(\mu_R^2)) \delta(1-z_1)\delta(1-z_2) \,.
\end{eqnarray}
If $\Phi^{~I}_d(\hat a_s,q^2,\mu^2,z_1,z_2,\ep)$ contains the correct poles
to cancel the poles
coming from $\hat F^I$,$Z^I$ and $\Gamma_{II}$ in order to
make $\Delta^{\rm sv}_{~d,I}$ finite, then the constants $\overline A^{~I}$ 
have to satisfy
\begin{eqnarray}
\overline A^{~I}=-A^I \,.
\end{eqnarray}
Using the above relation, the solution to the 
RG equation for $\overline G^{~I}_d
\left(\hat a_s,q^2/\mu_R^2,\mu_R^2/\mu^2,z_1,z_2,\ep\right)$ 
is found to be 
\begin{eqnarray}
\overline G^{~I}_d\left(\hat a_s,{q^2 \over \mu_R^2},
{\mu_R^2 \over \mu^2},z_1,z_2,\ep\right)
&=&\overline G^{~I}_d \left(a_s(\mu_R^2),{q^2 \over \mu_R^2},z_1,z_2,\ep\right)
\nonumber\\[2ex]
&=&\overline G^{~I}_d\left(a_s(q^2),1,z_1,z_2,\ep\right)
\nonumber\\[2ex]
&&- \delta(1-z_1)\delta(1-z_2) \int_{q^2/\mu_R^2}^1
{d\lambda^2 \over \lambda^2} A^I\left(a_s(\lambda^2 \mu_R^2)\right) \,.
\end{eqnarray}
With these solutions, it is straightforward to solve the Sudakov
differential equations yielding
\begin{eqnarray}
\Phi^{~I}_d(\hat a_s,q^2,\mu^2,z_1,z_2,\ep) &=& 
\Phi^{~I}_d(\hat a_s,q^2 (1-z_1)(1-z_2),\mu^2,\ep)
\nonumber\\[2ex]
&=&\sum_{i=1}^\infty \hat a_s^i \left({q^2 (1-z_1)(1-z_2) 
\over \mu^2}\right)^{i {\ep \over 2}}\!\! S_{\ep}^i 
\left({(i~\ep)^2 \over 4(1-z_1) 
(1-z_2)} \right)
\hat \phi^{~I,(i)}_d(\ep)\,,
\label{softsol}
\end{eqnarray}
where 
\begin{eqnarray}
\hat \phi^{~I,(i)}_d(\ep)=
{1 \over i \ep} \Bigg[ \overline K^{~I,(i)}_d(\ep) 
+ \overline {G}^{~I,(i)}_d(\ep)\Bigg]\,.
\end{eqnarray}
The above solutions for $\Phi^I_d$ satisfy the fact that 
$\Delta^{\rm sv}_{d,I}$ are finite as $\ep \rightarrow 0$ (see eqn.(\ref{DYH})).  
The constants $\overline K^{~I,(i)}_d(\ep)$
are determined by expanding $\overline K^I_d$ in powers of the 
bare coupling constant $\hat a_s$ as follows
\begin{eqnarray}
\overline K^I_d\left(\hat a_s,{\mu_R^2\over \mu^2},z_1,z_2,\ep\right)
=\delta(1-z_1)\delta(1-z_2) \sum_{i=1}^\infty \hat a_s^i
\left({\mu_R^2 \over \mu^2}\right)^{i {\ep \over 2}}S^i_{\ep}~
\overline K^{~I,(i)}_d(\ep) \,,
\end{eqnarray}
and solving the RG equation for
$\overline K^I_d\left(\hat a_s,\mu_R^2/\mu^2,z_1,z_2,\ep\right)$.  
The constants $\overline K^{~I,(i)}_d(\ep)$ are identical to $\overline K^{~I,(i)}(\ep)$ given 
in \cite{Ravindran:2006cg}.
The constants $\overline {G}^{~I,(i)}_d(\ep)$ are related to the finite
boundary functions $\overline G^I_d(a_s(q^2),1,z_1,z_2,\ep)$. 
We define the $\overline {\cal G}_{d,i}^I(\ep)$ through the relation 
\begin{eqnarray}
\sum_{i=1}^\infty \hat a_s^i 
\left( {q^2 (1-z_1)(1-z_2) \over \mu^2}\right)^{i{\ep \over 2}} 
S^i_{\ep}
\overline G_d^{~I,(i)}(\ep)
&=&
\sum_{i=1}^\infty a_s^i\left(q^2 (1-z_1)(1-z_2)\right) 
\overline {\cal G}^{~I}_{d,i}(\ep)
\label{Gbar1}
\end{eqnarray}

The $z_1,z_2$ independent constants
$\overline {\cal G}^{~I}_{d,i}(\ep)$ 
are obtained by demanding the finiteness of $\Delta^{\rm sv}_{~d,I}$ given in
eqn.(\ref{master}).   
Without setting $\ep=0$ in eqn.(\ref{master}), we expand 
$\Delta^{\rm sv}_{~d,I}$ as
\begin{eqnarray}
\Delta^{\rm sv}_{~d,I}(z_1,z_2,q^2,\mu_R^2,\mu_F^2,\ep)=\sum_{i=0}^\infty a_s^i(\mu_R^2)
\Delta^{\rm sv,(i)}_{~d,I}(z_1,z_2,q^2,\mu_R^2,\mu_F^2,\ep)\,.
\end{eqnarray}
Using the above expansion and eqn.(\ref{DYH}) we determine 
these constants by comparing the pole as well as non-pole terms
of the form factors, the mass factorisation kernels and the coefficient
functions $\Delta^{\rm sv,(i-1)}_{~d,I}$ expanded in powers of $\ep$ to
the desired accuracy.  
Since the $G^{~I}_d(\ep)$ in the form factors are found to satisfy 
a specific structure in terms of $f^I$,$\beta_i$
\cite{Ravindran:2006cg},  
we find that the constants $\overline {\cal G}^{~I}_{d,i}(\ep)$
also satisfy the following expansions containing these constants.
\begin{eqnarray}
\overline {\cal G}^{~I}_{d,1}(\ep)&=&-f_1^I+
\sum_{k=1}^\infty \ep^k \overline {\cal G}^{~I,(k)}_{d,1}\,,
\nonumber\\[2ex]
\overline {\cal G}^{~I}_{d,2}(\ep)&=&-f_2^I
-2 \beta_0 \overline{\cal G}_{d,1}^{~I,(1)}
+\sum_{k=1}^\infty\ep^k  \overline {\cal G}^{~I,(k)}_{d,2}\,,
\nonumber\\[2ex]
\overline {\cal G}^{~I}_{d,3}(\ep)&=&-f_3^I
-2 \beta_1 \overline{\cal G}_{d,1}^{~I,(1)}
-2 \beta_0 \left(\overline{\cal G}_{d,2}^{~I,(1)}
+2 \beta_0 \overline{\cal G}_{d,1}^{~I,(2)}\right)
+\sum_{k=1}^\infty \ep^k \overline {\cal G}^{~I,(k)}_{d,3}\,,
\nonumber \\[2ex]
\overline {\cal G}^{~I}_{d,4}(\ep)&=&
-f_4^I
-2 \beta_2 \overline {\cal G}^{~I,(1)}_{d,1}
-2 \beta_1 \Big( \overline {\cal G}^{~I,(1)}_{d,2} 
+ 4 \beta_0 \overline {\cal G}^{~I,(2)}_{d,1} \Big)\,,
\nonumber\\[2ex]
&&-2 \beta_0 \Big(\overline {\cal G}^{~I,(1)}_{d,3}
+2 \beta_0 \overline {\cal G}^{~I,(2)}_{d,2}
                         +4 \beta_0^2 \overline {\cal G}^{~I,(3)}_{d,1}\Big)
        +\sum_{k=1}^\infty \ep^k \overline {\cal G}^{~I,(k)}_{d,4}\,.
\label{OGI}
\end{eqnarray}

Now that we have a better understanding
\cite{Ravindran:2004mb} of the
structure of even the single pole terms of the form factors,  
we can predict all the poles {\it including the
single pole} of the soft distribution function from those of the form factors,
the renormalisation constants and the mass factorisation kernels.
The coefficients of the single poles are proportional to the constants 
$-f^I_i$ which are not only process independent but also maximally non-abelian.
The $\ep$ dependent terms in $\overline {\cal G}^{~I}_d(\ep)$ can be obtained
from the fixed order (in $a_s$) computations of cross sections
and the finite parts of the form factors.  At the moment,
we know $\overline {\cal G}^{~I}_{d,1}(\ep)$ to all orders in $\ep$,
$\overline {\cal G}^{~I}_{d,2}(\ep)$ to order $\ep$ and $\overline 
{\cal G}^{~I}_{d,3}(\ep)$ to order ${\ep}^0$.
The lowest order term $\overline {\cal G}^{~I}_{d,1}(\ep)$
is known to all orders in $\ep$ from  
the exact fixed-order soft contribution at order $a_s$.  
The next-to-leading order in $a_s$ (lowest order) computations 
of the total cross sections for DY and Higgs production 
determine the constants $\overline {\cal G}^{~I}_{d,1}(\ep)$ 
and the results reveal that they are maximally non-abelian to all 
orders in $\ep$. 
One can similarly determine the $\ep$ dependent parts 
of soft cross sections beyond the order $a_s$.  
The easier method is to use total cross sections that are known upto
NNLO level.  
We can easily extract $\overline {\cal G}^{~I}_{d,2}(\ep)$ upto order
$\ep$ by using the fact that these constants  
are independent of $z_j$ ($j=1,2$) and the differential cross sections satisfy
the relations 
\begin{eqnarray}
\int_0^1 dx_1^0 \int_0^1 dx_2^0  
\left(x_1^0 x_2^0\right)^{N-1}
(x_1^0+x_2^0) {d \sigma^I \over d x_F}
=
\int_0^1 
dx_1^0 \int_0^1 
dx_2^0 \left(x_1^0 x_2^0\right)^{N-1}
{d \sigma^I \over d Y}
=\int_0^1 d\tau~ \tau^{N-1} ~\sigma^I\,,
\label{iden}
\end{eqnarray}
while the $\sigma^I$ are known for both DY and Higgs production upto NNLO level
\cite{Matsuura:1987wt,Matsuura:1988sm,Hamberg:1990np,Harlander:2001is,
Catani:2001ic,Catani:2003zt,Harlander:2002wh,Anastasiou:2002yz,Ravindran:2003um,
Harlander:2003ai,Ravindran:2004mb}.
An alternative method is to take $N\rightarrow \infty$ on both sides of 
eqn.(\ref{iden}).
In this limit,  we find the following useful relation between
the constants $\hat \phi^{I,(i)}_d(\ep)$ that appear in
eqn.(\ref{softsol}) and $\hat \phi^{I,(i)}(\ep)$ 
that contribute to the soft distribution function of the total cross section:
\begin{eqnarray}
\hat \phi^{I,(i)}_d(\ep)=
{\Gamma(1+i~\ep) \over \Gamma^2\left(1+i{\ep \over 2}\right)}
\hat \phi^{I,(i)}(\ep).
\end{eqnarray}
Both the methods give
\begin{eqnarray}
\overline{\cal G}^{~I,(1)}_{d,1}
&=&C_I~ \Big(- \zeta_2\Big) \,,
\nonumber\\[2ex]
\overline{\cal G}^{~I,(2)}_{d,1}
&=& C_I~ \Bigg({1 \over 3}  \zeta_3\Bigg)\,,
\nonumber\\[2ex]
\overline{\cal G}^{~I,(3)}_{d,1}
&=& C_I~ \Bigg({1 \over 80}  \zeta_2^2\Bigg)\,,
\nonumber\\[2ex]
\overline{\cal G}^{~I,(1)}_{d,2}
&=& C_I C_A~ \Bigg({2428 \over 81} -{67 \over 3} \zeta_2
              -4 \zeta_2^2 -{44 \over 3} \zeta_3\Bigg)
\nonumber\\[2ex]
&&             +C_I n_f~ \Bigg(-{328 \over 81} + {10 \over 3} \zeta_2
                +{8 \over 3} \zeta_3 \Bigg)\,,
\label{DIi}
\end{eqnarray}
where $C_I=C_F$ for $I=q,b$ and $C_I=C_A$ for $I=g$.
Interestingly these constants $\overline {\cal G}^{~I}_{d,i}(\ep)$ turn out to
be maximally non-abelian.  That is, they satisfy
\begin{eqnarray}
\overline {\cal G}^{~q}_{d,i}(\ep) = 
\overline {\cal G}^{~b}_{d,i}(\ep) = 
{C_F \over C_A}~ \overline {\cal G}^{~g}_{d,i}(\ep) \,.
\end{eqnarray}
This implies that the soft distribution functions for the differential
cross sections satisfy
\begin{eqnarray}
 \Phi^{~q}_{d}\left(\hat a_s,q^2,  \mu^2,z_1,z_2,\ep\right) = 
\Phi^{~b}_{d}\left(\hat a_s,q^2 ,\mu^2,z_1,z_2,\ep\right) = 
{C_F \over C_A}~ \Phi^{~g}_{d}\left(\hat a_s,q^2 ,\mu^2,z_1,z_2,\ep\right)\,,
\end{eqnarray}
{\it upto order $a_s^2$},  
similar to the soft distributions that appear in the total cross sections.
We expect that this property will hold to all orders in perturbation theory
because of the fact that it originates entirely from the soft part of the
differential cross sections.

The threshold corrections dominate when the partonic scaling 
variables $z_1$ and $z_2$ approach their kinematic limit, which is unity,  
through the distributions $\delta(1-z_j)$ and ${\cal D}_i(z_j)$ with $j=1,2$.  
Resummations of threshold enhanced contributions are usually
done in Mellin $N$ space which has been a successful approach. See  
\cite{Sterman:1986aj, Catani:1989ne, Contopanagos:1996nh, Eynck:2003fn}
for resummation of total cross sections.
We show in the following how the soft distribution functions 
$\Phi^I_d(\hat a_s,q^2,\mu^2,z_1,z_2,\ep)$
capture all the features of the $N$ space resummation approach. 
The exponents of the $z_j$ (with $j=1,2$) space resummed cross sections
get contributions from the form factors 
through $\delta(1-z_1) \delta(1-z_2)$ terms
and from the soft distribution
functions through $\delta(1-z_j)$ 
as well as the distributions ${\cal D}_i(z_j)$.  
We can rewrite the soft distribution function as
\begin{eqnarray}
\Phi^{~I}_d(\hat a_s,q^2,\mu^2,z_1,z_2,\ep)&=&
{1 \over 2}\delta(1-z_2) \Bigg( {1 \over 1-z_1} \Bigg\{
\int_{\mu_R^2}^{q^2 (1-z_1)} {d \lambda^2 \over \lambda^2}
A_I \left(a_s(\lambda^2)\right) 
\nonumber\\[2ex]
&&\hspace{1.2cm}+ \overline G^{~I}_d \left(
a_s\left(q^2 (1-z_1)\right),\ep\right)\Bigg\} \Bigg)_+
\nonumber\\[2ex]
&&+q^2 {d \over dq^2} \Bigg[
\Bigg({1 \over 4 (1-z_1) (1-z_2)}
\Bigg\{\int_{\mu_R^2}^{q^2 (1-z_1)(1-z_2)} {d \lambda^2 \over \lambda^2} 
A^I\left(a_s(\lambda^2)\right)
\nonumber\\[2ex]
&&\hspace{1.2cm}+\overline G^I_d\left(a_s\left(q^2 (1-z_1)(1-z_2)\right),\ep\right)\Bigg\}\Bigg)_+\Bigg]
\nonumber\\[2ex]
&&+{1 \over 2} \delta(1-z_1)\delta(1-z_2)~~~ \sum_{i=1}^\infty \hat a_s^i
\left({q^2 \over \mu^2}\right)^{i {\ep \over 2}}
S_{\ep}^i~
\hat \phi^{~I,(i)}_d(\ep)
\nonumber\\[2ex]
&&+{1 \over 2} \delta(1-z_2)
\left({1 \over 1-z_1}\right)_+ ~~~\sum_{i=1}^\infty
\hat a_s^i \left({\mu_R^2 \over \mu^2}\right)^{i {\ep \over 2}}
S_{\ep}^i~
\overline K^{~I,(i)}(\ep)
\nonumber\\[2ex]
&&+ (z_1 \leftrightarrow z_2)\,,
\label{resum}
\end{eqnarray}
where
\begin{eqnarray}
\overline G^{~I}_d\left(a_s \left(q^2 g(z_1,z_2) \right),\ep\right)
&=& \sum_{i=1}^\infty \hat a_s^i
\left({q^2 g(z_1,z_2) \over \mu^2}\right)^{i{\ep \over 2}}
S_{\ep}^i
\overline G^{~I,(i)}_d(\ep)\,.
\label{Gbar2}
\end{eqnarray}
The third term in eqn.(\ref{resum}) contains the correct poles
in $\ep$ to cancel those coming from the form factors as well as the
$\delta(1-z_j)$ parts of the mass factorisation kernels.  
The fourth term contains only poles that cancel 
against the ${\cal D}_0(z_j)$ parts of the mass factorisation kernels.
The remaining finite terms as $\ep$ becomes zero in the first three terms 
contribute to the soft-plus-virtual parts of the differential cross sections.  
Hence, adding the eqn.(\ref{resum}) to the renormalised form factors and the
mass factorisation kernels, performing the coupling constant renormalisation,
and then finally taking the double Mellin moment in $N_1,N_2$, 
we get the resummed result analogous to the threshold resummation formula 
that one obtains
for the total inclusive cross sections(see 
\cite{Sterman:1986aj, Catani:1989ne, Contopanagos:1996nh, Eynck:2003fn})
when $\ep \rightarrow 0$. A similar result for the resummed rapidity
distribution scheme can be found 
in \cite{Catani:1989ne}. 

In the double Mellin space ($N_1,N_2$) the threshold enhanced differential 
cross section will be proportional to
\begin{eqnarray}
\exp \Big[ 2 
\int_0^1 dz_1 z_1^{N_1-1}
\int_0^1 dz_2 z_2^{N_2-1}
\Phi^I_{d,{\rm finite}}(\hat a_s, q^2,\mu^2,z_1,z_2)\Big]\,.
\label{exp}
\end{eqnarray}

Similar to the soft distribution functions 
$\Phi^{~I}_P(\hat a_s,q^2,\mu^2,z,\ep)$ that enter in 
DY and Higgs production
\cite{Ravindran:2005vv}, 
the present $\Phi^{~I}_d(\hat a_s,q^2,\mu^2,z_1,z_2,\ep)$
are also maximally non-abelian.  
Using the resummed result given in eqn.(\ref{master}), and the
exponents $g_i^{~I}(\ep)$(see \cite{Moch:2005tm}), $\overline {\cal G}_{d,i}^{~I}(\ep)$, 
we can obtain the higher order soft-plus-virtual 
contributions to the differential cross sections.  The available
exponents are 
\begin{eqnarray}
&g_1^{~I,j}~,~~~
\overline {\cal G}_{d,1}^{~I,(j)} \hspace{1.5cm} 
{\rm for}\hspace{1.5cm}  j={\rm all} \quad \,,
\nonumber\\[2ex]
&g_2^{~I,j}~,~~~
\overline {\cal G}_{d,2}^{~I,(j)} \hspace{1.5cm} 
{\rm for} \hspace{1.5cm} j=0,1 \quad \,,
\nonumber\\[2ex]
&g_3^{~I,j}~,~~~
\overline {\cal G}_{d,3}^{~I,(j)} \hspace{1.5cm} 
{\rm for} \hspace{1.5cm} j=0 \quad \,,
\nonumber
\end{eqnarray}
in addition to the known $\beta_i~(i=0,1,2,3)$, the constants in the splitting
functions $A^I_i,~ B^I_i~~(i=1,2,3)$, the maximally non-abelian constants
$f^I_i~(i=1,2,3)$ and the anomalous dimensions $\gamma^b_i(i=0,1,2,3)$.  For
$I=q,g$, the constants $g^{q,j}_2$ and $g^{g,j}_2$ are known for $j=2,3$ also
(see \cite{Moch:2005id}).  Using the resummed expression given in 
eqn.(\ref{master}) and the known exponents, we present here the 
results for $\Delta_{d,I}^{\rm sv,(i)}$ for DY and Higgs production.  
Using our approach we have first reproduced the Drell-Yan coefficient 
$\Delta_{d,q}^{\rm sv,(i)}$, known upto NNLO ($i=0,1,2$) \cite{Rijken:1994sh}.
We then obtain $\Delta_{d,g}^{\rm sv,(i)}$ and $\Delta_{d,b}^{\rm sv,(i)}$ 
for the Higgs production upto NNLO ($i=1,2$).  For N${}^3$LO for $I=q,b,g$, 
{\it a partial result} $\Delta_{d,I}^{\rm sv,(3)}$, i.e., a result without 
the $\delta(1-z_1)\delta(1-z_2)$ part can be computed from our 
formula given in eqn.(\ref{master}).  
The coefficient of $\delta(1-z_1)\delta(1-z_2)$ part depends on
still unknown constants $\overline {\cal G}^{I,(2)}_2,g^{I,1}_3,\overline 
{\cal G}^{I,(1)}_3$.
We can also obtain a result  
to N${}^4$LO order where we can predict {\it partial} soft-plus-virtual
contributions containing everything except 
the terms in ${\cal D}_0(z_i) \delta(1-z_j),{\cal D}_0(z_i) {\cal D}_0(z_j),
{\cal D}_1(z_i) \delta(1-z_j)$ and $\delta(1-z_1) \delta(1-z_2)$ for
the Drell-Yan N${}^4$LO coefficient $\Delta_{d,q}^{\rm sv,(4)}$, 
the gluon fusion to Higgs N${}^4$LO coefficient $\Delta_{d,g}^{\rm sv,(4)}$ and
the bottom quark annihilation to Higgs boson N${}^4$LO coefficient
$\Delta_{d,b}^{\rm sv,(4)}$.  
The results are presented in  
the Appendix B for $\mu_R^2=\mu_F^2=q^2$.   
The convolutions of distributions of the 
form ${\cal D}_l(z_j) \otimes {\cal D}_m(z_j)$ for any arbitrary $l,m$ 
can be done using the general formula given in \cite{Ravindran:2006cg}.
Using these convolutions it is straightforward to calculate 
$\Delta_{d,I}^{\rm sv,(i)}$ for $i=1,...,4$ for both DY ($I=q$) and
Higgs ($I=g,b$) production. An alternative derivation of the NNLO DY
soft-plus-virtual terms in the DIS renormalisation
scheme can be found in \cite{ac}, where they calculated di-lepton production
cross sections at fixed target energies.  

\begin{figure}[htb]
\vspace{1mm}
\centerline{
\epsfig{file=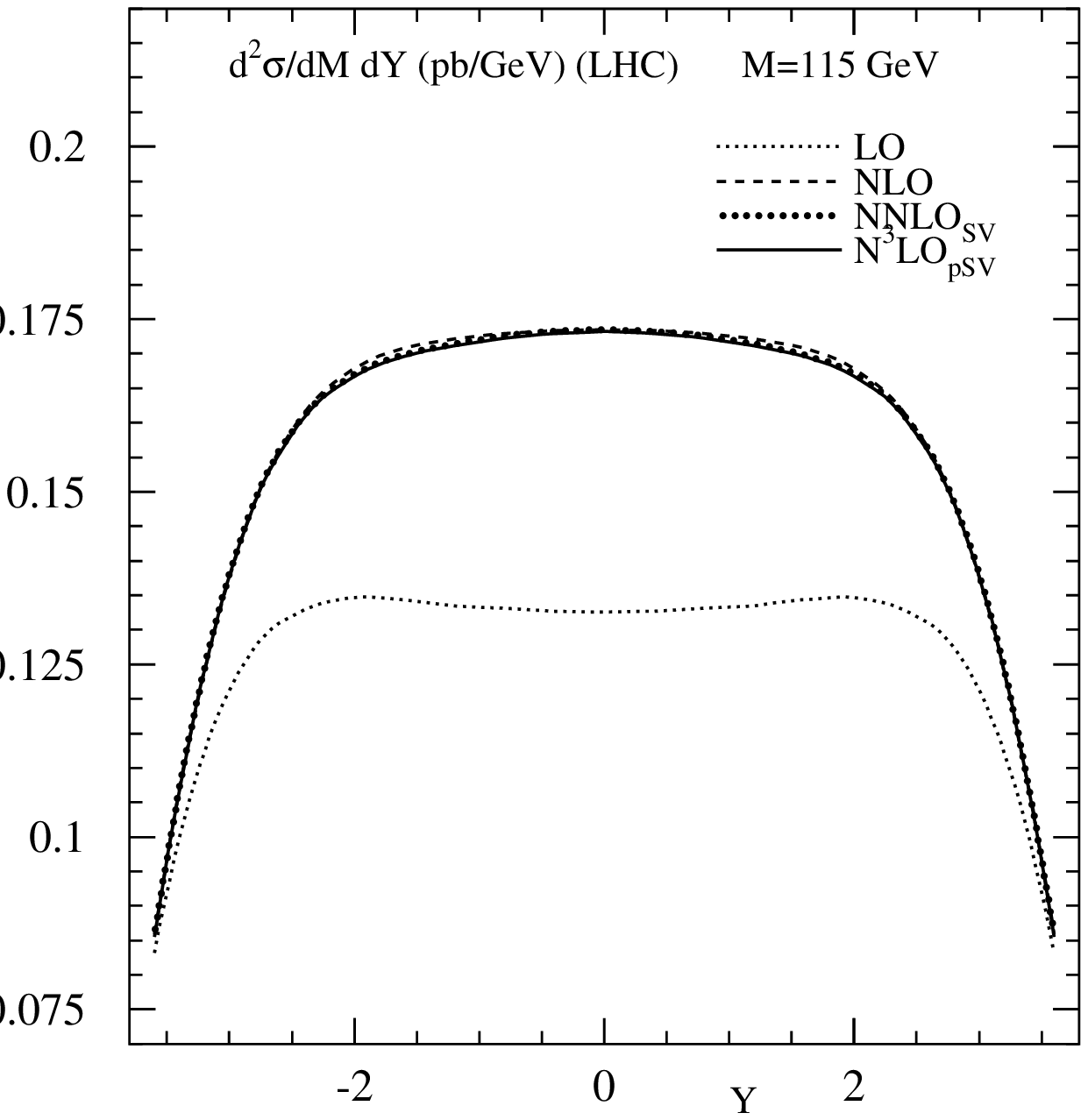,width=8cm,height=10cm,angle=0}
\epsfig{file=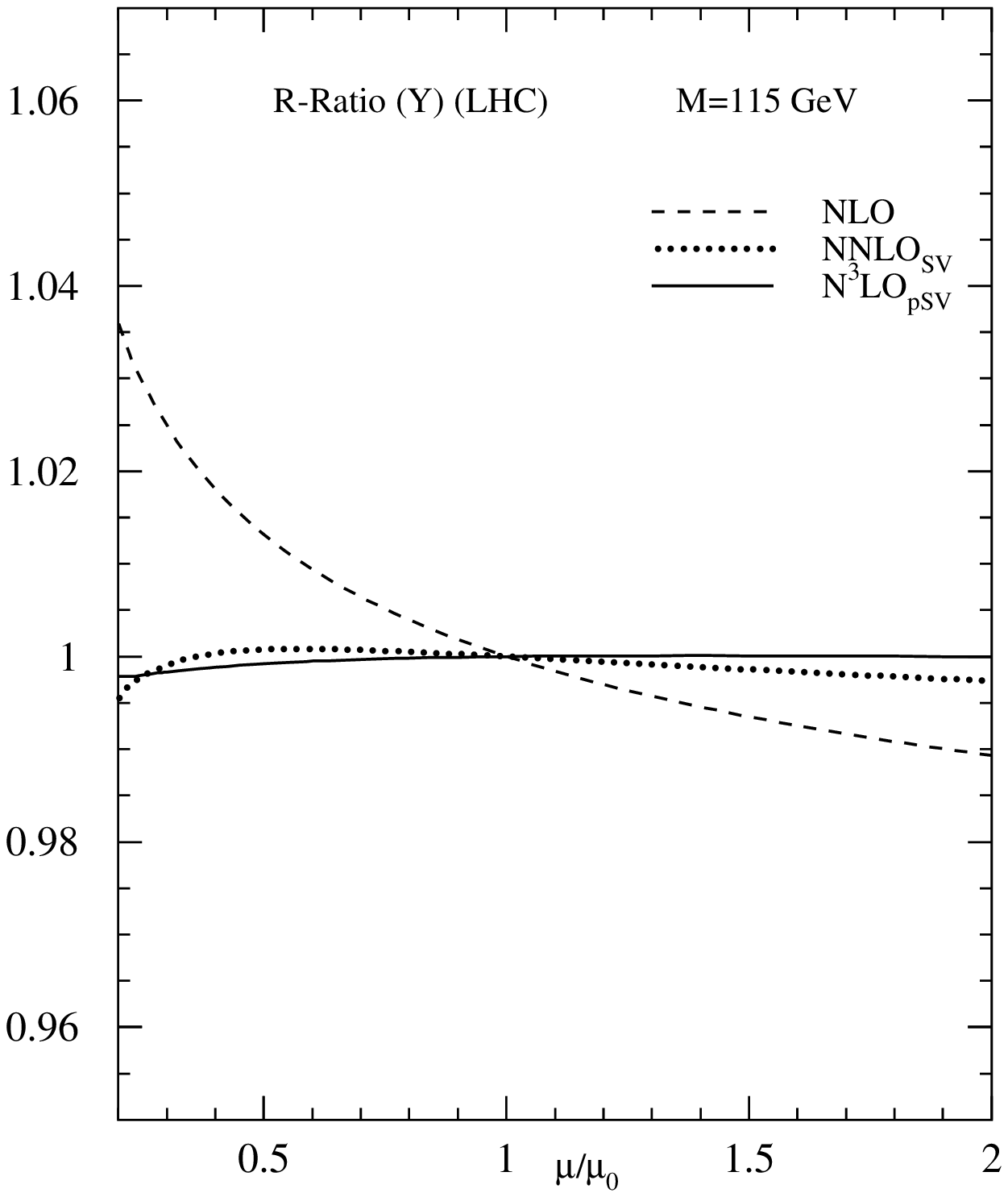,width=8cm,height=9.3cm,angle=0}
}
\caption{
Rapidity distributions for DY production at the LHC,
and their $\mu_R$ scale dependence(with $\mu_F^2=q^2=M^2_{l^+l^-}$). 
Here we denote $M=M_{l^+l^-}$.  The abbreviation "pSV" 
means partial-soft-plus-virtual.
}

\end{figure}
\begin{figure}[htb]
\vspace{1mm}
\centerline{
\epsfig{file=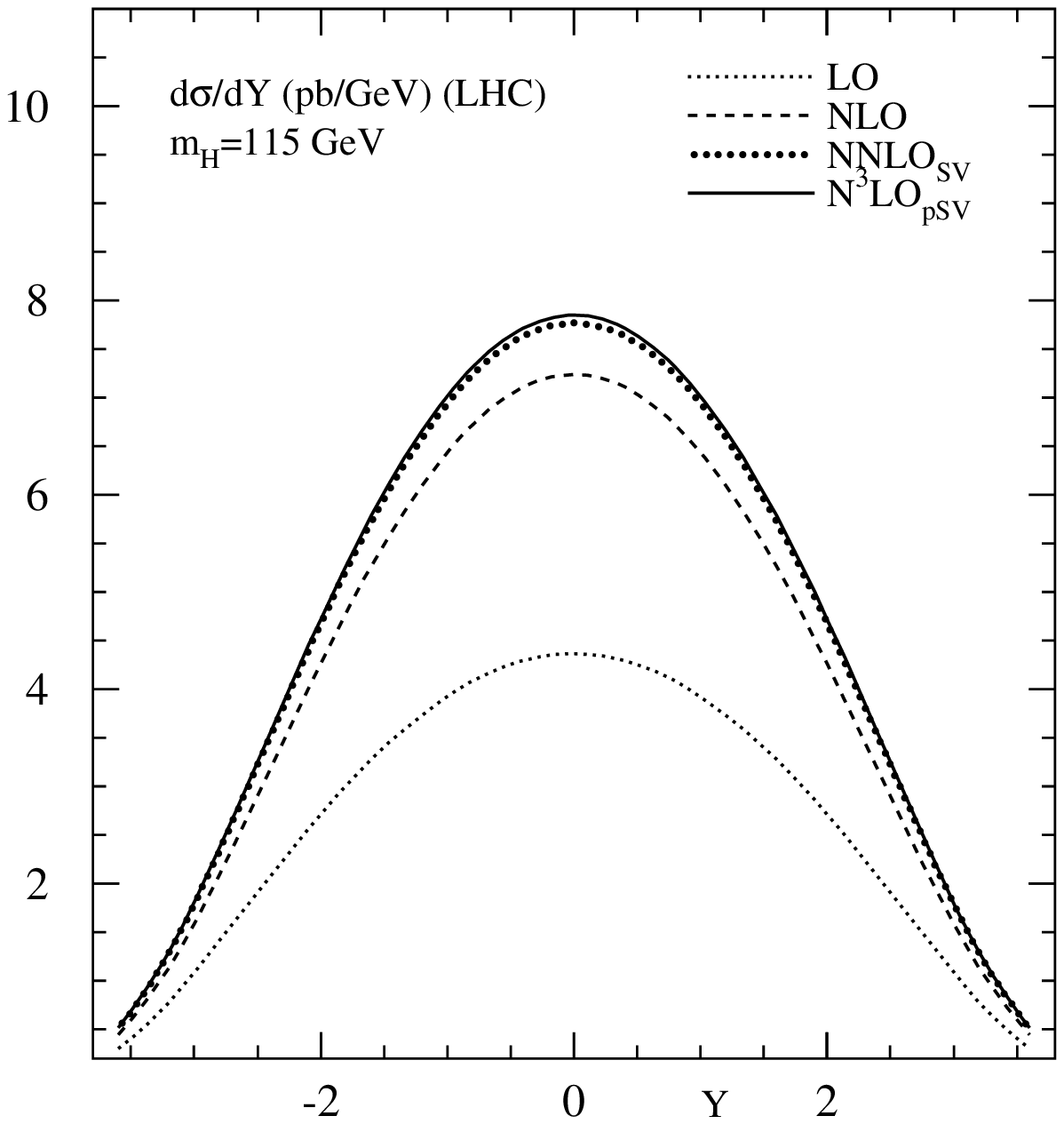,width=8cm,height=10cm,angle=0}
\epsfig{file=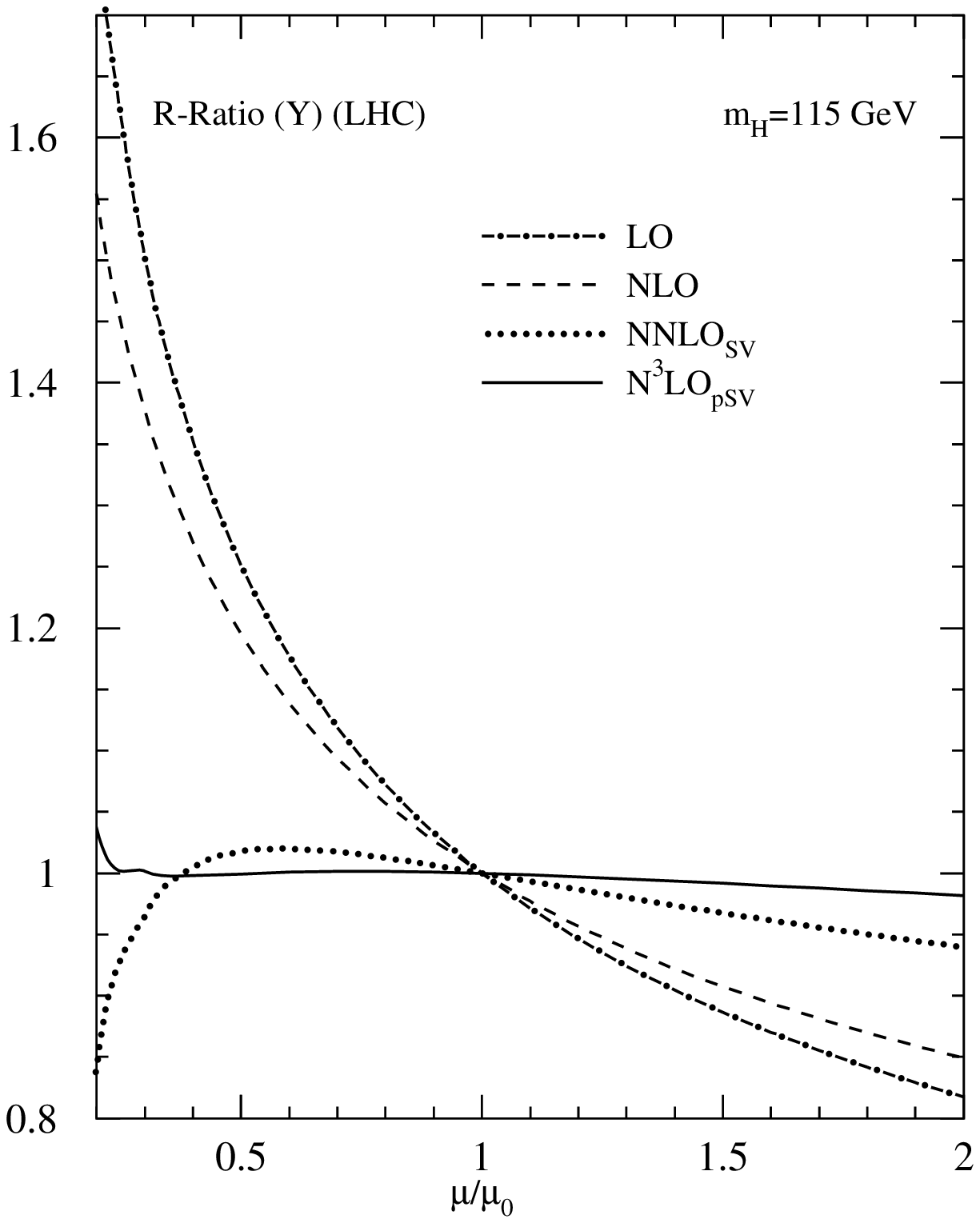,width=8cm,height=9.3cm,angle=0}
}
\caption{
Rapidity distributions for Higgs production through gluon fusion at LHC,
and their $\mu_R$ scale dependence for the Higgs boson of mass 
$m_H=115~{\rm GeV}$ with $\mu_F=m_H$. 
}
\end{figure}

The differential cross sections for $I=q$ can be expanded in powers of
the strong coupling constant as 
\begin{eqnarray}
{d \sigma^I\over  dY} &=& \sum_{i=0}^\infty a_s^i 
~{d \sigma^{I,(i)} \over dY} \,.
\end{eqnarray}
We split the partonic cross section into hard and sv parts:
\begin{eqnarray}
{d \sigma^{I,(i)} \over dY} = {d \sigma^{{\rm hard},I,(i)} \over dY}
+{d \sigma^{\rm sv,I,(i)} \over dY}\,.
\end{eqnarray}
\begin{eqnarray}
2 S~ {d^2 \sigma^{{\rm hard},q,(i)}\over dq^2 dY} &=&
\sum_{q} {\cal G}_{SM,q} 
\Bigg( D^{SM,(i)}_{q \overline q}(\xo,\xt,\mu_F^2)
      +D^{SM,(i)}_{q g}(\xo,\xt,\mu_F^2)
\nonumber\\[2ex]&&
      +D^{SM,(i)}_{g q}(\xo,\xt,\mu_F^2)\Bigg)
\nonumber\\[2ex]
2 S~ {d \sigma^{{\rm hard},g,(i)}\over dY} &=&
{\cal G}_H 
D^{H,(i)}_{gg}(\xo,\xt,\mu_F^2)
+\sum_{q}
{\cal G}_H 
      \Bigg(D^{H,(i)}_{q g}(\xo,\xt,\mu_F^2)
\nonumber\\[2ex]&&
      +D^{H,(i)}_{g q}(\xo,\xt,\mu_F^2)
      +D^{H,(i)}_{q \overline q}(\xo,\xt,\mu_F^2)\Bigg)
\end{eqnarray}

\begin{table}[hbp]
\begin{center}
\begin{tabular}{|l|l|l|l|l|}
\hline
$Y$ & LO & NLO & N${}^2$LO${}_{\rm SV}$ & N${}^3$LO${}_{\rm pSV}$\\
\hline
\hline
  0.   &0.1326  &0.1735  &0.1734  &0.1732\\
  0.4  &0.1327  &0.1733  &0.1732  &0.1729\\
  0.8  &0.1330  &0.1729  &0.1725  &0.1722\\
  1.2  &0.1335  &0.1721  &0.1714  &0.1711\\
  1.6  &0.1343  &0.1707  &0.1698  &0.1695\\
  2.   &0.1346  &0.1678  &0.1670  &0.1667\\
  2.4  &0.1328  &0.1614  &0.1610  &0.1607\\
  2.8  &0.1273  &0.1483  &0.1485  &0.1482\\
  3.2  &0.1123  &0.1240  &0.1244  &0.1241\\
  3.6  &0.0832  &0.0858  &0.0857  &0.0855\\
\hline
\end{tabular}
\end{center}
\caption{ Values for $d^2\sigma/dMdY$ (M=115 GeV) in pb/GeV 
at fixed values of $Y$ which are plotted in figure 1.} 
\end{table}

The SM coefficients  $D^{SM,(i)}_{ab}(\xo,\xt,\mu_F^2)$ can be found in
\cite{Rijken:1994sh,Mathews:2004xp},
and the $D^{H,(i)}_{ab}(\xo,\xt,\mu_F^2)$ for the Higgs are given in
Appendix A.
The soft-plus-virtual part of the DY partonic cross section can be expressed as
\begin{eqnarray}
2 S~ {d^2 \sigma^{\rm sv,q,(i)} \over dq^2 dY} 
= \sum_{a,b=q,\overline q}{\cal G}_{SM,q}
\int_0^1~ dx_1~ \int_0^1~ dx_1~ {\cal H}^q_{ab}(x_1,x_2,\mu_F^2)~ 
\nonumber\\[2ex]
\times \int_0^1 dz_1 \int_0^1 dz_2 ~\delta(x_1^0-x_1 z_1) 
~\delta(x_2^0-x_2 z_2)~ \Delta_{Y,q}^{\rm sv,(i)}
(z_1,z_2,q^2,\mu_F^2,\mu_R^2)\,,
\end{eqnarray}
and for the Higgs partonic cross section we have 
\begin{eqnarray}
2 S~ {d \sigma^{\rm sv,g,(i)} \over dY} 
={\cal G}_{H} 
\int_0^1~ dx_1~ \int_0^1 ~dx_2 ~ {\cal H}^g_{gg}(x_1,x_2,\mu_F^2)~ 
\nonumber\\[2ex]
\times \int_0^1 dz_1 \int_0^1 dz_2 ~\delta(x_1^0-x_1 z_1) 
~\delta(x_2^0-x_2 z_2)~\Delta_{Y,g}^{\rm sv,(i)}
(z_1,z_2,m_H^2,\mu_F^2,\mu_R^2)\,.
\end{eqnarray}
The coefficients $\Delta_{Y,ab}^{\rm sv,(i)}(z_1,z_2,q^2,\mu_F^2,\mu_R^2)$ are 
presented in the Appendix B, where we use the normalisation 
$\Delta_{Y,ab}^{\rm sv,(0)}(z_1,z_2,q^2,\mu_F^2,\mu_R^2)=
\delta(1-z_1)\delta(1-z_2)$.
The constants ${\cal G}_{SM,q},{\cal G}_H$ are given by
\begin{eqnarray}
{\cal G}_{SM,q}&=&{4 \alpha^2 \over 3 q^2} \Bigg[Q_q^2
-{2 q^2 (q^2-M_Z^2) \over
\left((q^2-M_Z^2)^2+M_Z^2 \Gamma_Z^2\right) c_w^2 s_w^2}
Q_q g_e^V g_q^V
\nonumber\\[2ex]
&&+{q^4 \over  \left((q^2-M_Z^2)^2+M_Z^2 \Gamma_Z^2\right)
c_w^4 s_w^4}\Big((g_e^V)^2+(g_e^A)^2\Big)\Big((g_q^V)^2+(g_q^A)^2\Big)
\Bigg]\,,
\nonumber\\[2ex]
\label{ew}
\\[2ex]
{\cal G}_{H}&=&{\pi m_H^2 G_B^2 \over 4 (N^2-1)}\,.
\label{gr}
\end{eqnarray}
The $x_F$ differential cross sections can be obtained 
from the $Y$ differential cross sections by 
replacing $D_{ab}^{I,(i)}(\xo,\xt,\mu_F^2)$ 
by $C_{ab}^{I,(i)}(\xo,\xt,\mu_F^2)$.  
For the sv part we identify 
$\Delta_{Y,I}^{\rm sv,(i)}=
(x_1^0+x_2^0)^{-1}  \Delta_{x_F,I}^{\rm sv,(i)}$
with the replacement of $D_{ab}^{I,(i)}(\xo,\xt,\mu_F^2)$ 
by $C_{ab}^{I,(i)}(\xo,\xt,\mu_F^2)$  
in the right-hand-side.
The electro-weak constants appearing in eqns.(\ref{ew},\ref{gr}) 
can be found in 
\cite{Mathews:2004xp,Ravindran:2003um}.

\begin{table}[hbp]
\begin{center}
\begin{tabular}{|l|l|l|l|l|}
\hline
$Y$ & LO & NLO & N${}^2$LO${}_{\rm SV}$ & N${}^3$LO${}_{\rm pSV}$\\
\hline
\hline
  0.   &4.366  &7.236  &7.765  &7.848\\
  0.4  &4.294  &7.106  &7.632  &7.713\\
  0.8  &4.084  &6.722  &7.236  &7.313\\
  1.2  &3.731  &6.091  &6.585  &6.655\\
  1.6  &3.278  &5.264  &5.724  &5.784\\
  2.   &2.715  &4.265  &4.675  &4.724\\
  2.4  &2.068  &3.173  &3.512  &3.549\\
  2.8  &1.410  &2.087  &2.337  &2.361\\
  3.2  &0.779  &1.135  &1.285  &1.298\\
  3.6  &0.300  &0.448  &0.511  &0.516\\
\hline
\end{tabular}
\end{center}
\caption{ Values for $d\sigma/dY$ (for $M_H=$ 115 GeV) in pb/GeV 
at fixed values of $Y$ which are plotted in figure 2.}
\end{table}

For our numerical results we choose the center-of-mass energy to
be $\sqrt{S}=$14 TeV for the LHC.
The standard model parameters that enter our computation
are the Fermi constant $G_F=4541.68$ pb, the $Z$ boson mass $M_Z=91.1876$ 
GeV and top quark mass $m_t=173.4$ GeV.  The strong coupling constant
$\alpha_s(\mu_R^2)$ is evolved using the 4-loop RG equations
depending on the order in which the cross section is evaluated.
We choose $\alpha^{\rm LO}_s(M_Z)=0.130$, $\alpha^{\rm NLO}_s(M_Z)=0.119$,
$\alpha^{\rm NNLO}_s(M_Z)=0.115$ and 
$\alpha^{{\rm N}{}^i{\rm LO}}_s(M_Z)=0.114$ for $i>2$.
We use MRST 2001 LO for leading order, MRST2001 NLO for NLO
and MRST 2002 NNLO for N${}^i$LO with $i>1$ \cite{Martin:2002dr,Martin:2001es}.
The impact of the soft-plus-virtual N${}^2$LO and the
partial soft-plus-virtual N${}^3$LO 
contributions to the DY rapidity differential cross section
at the LHC is presented in figure 1.
Note that we have not plotted the partial soft-plus-virtual 
N${}^4$LO contributions because there are no N${}^3$LO parton densities.
In the first plot we have shown the rapidity distribution 
in pb/GeV for a di-lepton mass of 115 GeV.  For LO and NLO, we used the exact
results which contain both the soft-plus-virtual as well as the regular hard
contributions.  For N${}^i$LO ($i=2,3$), we use only the soft-plus-virtual
results extracted from the resummed formula.  Here we have set
$\mu_F=\mu_R=115$ GeV.  We find that the inclusion of N${}^i$LO ($i=2,3$) 
terms only make small changes in the differential cross section which 
confirms the reliability of the perturbative approach.  
\begin{table}[hbp]
\begin{center}
\begin{tabular}{|l|l|l|l|l|}
\hline
$Y$ & LO & NLO & N${}^2$LO${}_{\rm SV}$ & N${}^3$LO${}_{\rm pSV}$\\
\hline
\hline
  0.   &4.085  &6.791  &7.293  &7.370\\
  0.4  &4.016  &6.668  &7.166  &7.241\\
  0.8  &3.815  &6.298  &6.785  &6.856\\
  1.2  &3.477  &5.693  &6.161  &6.225\\
  1.6  &3.042  &4.899  &5.334  &5.389\\
  2.   &2.505  &3.944  &4.330  &4.375\\
  2.4  &1.888  &2.906  &3.223  &3.256\\
  2.8  &1.266  &1.883  &2.112  &2.133\\
  3.2  &0.680  &0.997  &1.131  &1.143\\
  3.6  &0.250  &0.376  &0.429  &0.434\\
\hline
\end{tabular}
\end{center}
\caption{ Values for $d\sigma/dY$ (for $M_H=$ 120 GeV) in pb/GeV 
at fixed values of $Y$.}
\end{table}

In the second plot of fig.1 we have shown 
the scale variation of the rapidity distribution using the ratio:
\begin{eqnarray}
R^I\left({\mu_R^2}\right)=
\left({d\sigma^I\over dx}\left(\mu_R^2=q^2\right) \right)^{-1}
{d\sigma^I \over dx}\left(\mu_R^2\right) \,, 
\end{eqnarray} 
plotted as a function of 
$\mu/\mu_0=\mu_R/|q|$,
where we have fixed $\mu_F^2=q^2$.
It is clear from the second plot of fig.1 that the inclusion of 
N${}^i$LO ($i=2,3$) soft-plus-virtual contributions
further reduces the scale ambiguity.  

The impact of soft-plus-virtual parts N${}^2$LO and the
partial soft-plus-virtual N${}^3$LO 
contributions to Higgs production through gluon fusion 
at the LHC is presented in figure 2. We see that the inclusion of the
higher order terms does not make any appreciable change in the
magnitude of the rapidity distribution. Again this confirms
the reliability of the perturbation series. The second plot in figure 2
shows $R$ in eqn.48 as a function of 
$\mu/\mu_0=\mu_R/m_H$,
where we have fixed $\mu_F^2=m_H^2$,
and demonstrates that the inclusion of 
the higher order terms reduces the sensitivity to the choice of the scale.

We also present the numerical values of the rapidity 
distributions in figures 1 and 2 in Tables 1 and 2 respectively.
These numbers allow a more direct comparison with other theoretical
papers and are useful to the experimental groups working at the LHC. 

Previous calculations of differential distributions in NLO using the 
effective Lagrangian
(or the $m_t\rightarrow \infty$ approach) have been presented in
\cite{fgk}-\cite{Anastasiou:2002qz}. In the same approach the resummation 
of the logarithmically enhanced contributions 
have been carried out in \cite{boca} -\cite{field}. 
Our results agree exactly with the DY NLO results for rapidity
distributions in \cite{Hamberg:1990np}. We cannot compare the NLO Higgs
rapidity plots directly with those in \cite{rasm2} because there we used
$m_H= 120$ GeV and we had to impose a cut on the Higgs $p_t$.
However we have rerun our programs with $m_H=120$ GeV to allow a better
comparison with both the results in \cite{rasm2}. The number for the
Higgs rapidity distribution are given in Table 3. 
Our numbers are also consistent with the normalized Higgs boson rapidity 
distribution in fig. 1 in \cite{Anastasiou:2002qz}.
These checks indicate that everything is consistent with the NLO results.

We have compared our results for the Drell-Yan and Higgs rapidity 
distributions against the NNLO results published 
in \cite{Anastasiou:2003ds,Anastasiou:2003yy}.  
Our soft-plus-virtual NNLO  approximations agree very well with their exact 
NNLO results. Our partial soft-plus-virtual N${}^3$LO results are new and 
cannot be compared with any other calculation.  

To summarise, we have systematically studied the soft-plus-virtual
corrections to differential cross sections in rapidity for
DY and Higgs production through both gluon fusion and bottom quark 
annihilation.  The resummation of these corrections has been achieved  
using renormalisation group invariance, 
Sudakov resummation of scattering amplitudes and the factorisation
property of the hard scattering cross sections.  Our analytical results
are presented in Appendices A and B. It is now straightforward 
to obtain resummed threshold contributions to both $x_F$ and $Y$ 
rapidity distributions of di-lepton pairs in the DY process and of 
Higgs bosons in Higgs productions. This requires a 
double Mellin transform in the space of two variables $N_1$ and $N_2$,
see eqn.(40). 
Using our resummed results we have computed soft-plus-virtual 
differential cross sections at N${}^2$LO and 
partial-soft-plus-virtual differential cross sections  
in  N${}^3$LO.  Finally we have presented the numerical impact of these 
results on the rapidity differential cross sections.

{\bf Acknowledgments:}  
V. Ravindran and J. Smith would like to thank Prof. P. van Baal for
hospitality. 
The work of J. Smith has been partially supported by the
National Science Foundation, grant PHY-0098527. 
We all acknowledge support from the FOM under Grant No. L0102M and
discussions with E. Laenen, S. Moch, J. Vermaseren and A. Vogt.

\myappendix
\appendix
\section{Hart parts}
In this appendix, we list the
$C_{ab}^{H,(i)}(\xo,\xt)$ and $D_{ab}^{H,(i)}(\xo,\xt)$ 
that contribute to the hard parts of the cross sections.  We start by defining 
the following parton density combinations
\begin{eqnarray}
H_{q\overline q}(x_1,x_2,\mu_F^2)&=& 
f_q^{P_1}(x_1,\mu_F^2)~ f_{\overline q}^{P_2}(x_2,\mu_F^2)
+ f_{\overline q}^{P_1}(x_1,\mu_F^2)~ f_{q}^{P_2}(x_2,\mu_F^2)\,,
\nonumber\\[2ex]
H_{gq}(x_1,x_2,\mu_F^2)&=& 
f_g^{P_1}(x_1,\mu_F^2)~\left( f_{q}^{P_2}(x_2,\mu_F^2) 
  + f_{\overline q}^{P_2}(x_2,\mu_F^2)\right)\,,
\nonumber\\[2ex]
H_{qg}(x_1,x_2,\mu_F^2)&=& H_{gq}(x_2,x_1,\mu_F^2) \,,
\nonumber\\[2ex]
H_{gg}(x_1,x_2,\mu_F^2)&=& f_g^{P_1}(x_1,\mu_F^2)~ f_{g}^{P_2}(x_2,\mu_F^2)\,.
\end{eqnarray}
In terms of these combinations, we list the $C_{ab}^{H,(i)}(\xo,\xt)$ 
that appear in the hard parts of the $x_F$-differential cross sections 
\begin{eqnarray}
C_{gg}^{H,(0)}(\xo,\xt)&=& H_{gg}(\xo,\xt,\mu_F^2)\,,
\nonumber\\[2ex]
C_{gg}^{H,(1)}(\xo,\xt)&=& C_A \Bigg\{
\int_{\xo}^1 ~ dx_1 ~
          {H_{gg}(x_1,\xt,\mu_F^2) \over (\xo+\xt) }
\Bigg[\Bigg(
	  -4~{\xo^2 \over x_1^3}
	  +4~{\xo \over x_1^2}
	  -{8 \over x_1} 
	  +{4 \over \xo} 
      \Bigg) {\cal L}_{a_1}
\nonumber\\[2ex]&&
      +{4 \over (x_1-\xo)}~ {\cal L}_{c_1}
\Bigg]
+\int_{\xo}^1 ~dx_1~
	  {H_{gg,1}(x_1,\xt,\mu_F^2) \over (x_1-\xo)(\xo+\xt) }
	  \Bigg[
	  4 {\cal L}_{b_1}
	  \Bigg]
\nonumber\\[2ex]&&
+\int_{\xo}^1~dx_1~\int_{\xt}^1~dx_2~ 
       {H_{gg,1}(x_1,x_2,\mu_F^2) \over (x_1-\xo) (\xo+\xt)}
\Bigg[
	  -4~{\xt^2 \over x_2^3}
	  +4~{\xt \over x_2^2}
	  -{8 \over x_2} 
	  +{4 \over \xt} 
\Bigg] 
\nonumber\\[2ex]&&
            +\int_{\xo}~dx_1~\int_{\xt}^1~dx_2~
       {2 H_{gg,12}(x_1,x_2,\mu_F^2) \over (x_1-\xo)(x_2-\xt) (\xo+\xt)}
\nonumber\\[2ex]&&
            +{H_{gg}(\xo,\xt,\mu_F^2) \over  (\xo+\xt)} 
            \Bigg[2 {\cal L}(\xo,\xt) \log\left({q^2 \over \mu_F^2}\right)
       +\Big({\cal L}(\xo,\xt)\Big)^2
       +6 \zeta_2
\Bigg]
\nonumber\\[2ex]&&
     + \int_{\xo}^1~ dx_1~ \int_{\xt}^1~ dx_2~
	 {H_{gg}(x_1,x_2,\mu_F^2) \over 
	   (x_1+x_2)^3(x_1+\xt)(x_2+\xo)(\xo+\xt)}
\Bigg[
{1 \over x_1^3}\Bigg(
          4 x_2^3 \xo^2 \xt
\nonumber\\[2ex]&&
	  +4 x_2^2 (\xo^3 \xt + \xo^2 \xt^2)
	  + 4 x_2 ( \xo^3 \xt^2 + \xo^2 \xt^3)
	  \Bigg)
+{1 \over x_1^2} \Bigg(
         4 x_2^3 (\xo^2 - \xo \xt)
\nonumber\\[2ex]&&
	 +x_2^2 (4 \xo^3+12 \xo^2 \xt-4 \xo \xt^2)
	 +x_2 (16 \xo^3 \xt +12 \xo^2 \xt^2 -4 \xo \xt^3)
\nonumber\\[2ex]&&
	 +8 (\xo^3 \xt^2+\xo^2 \xt^3)
          \Bigg)
+{1 \over x_1} \Bigg(
         x_2^3 (8 \xt -4 \xo)
	 +x_2^2 (8 \xt^2 +8 \xo^2 -8 \xo \xt)
\nonumber\\[2ex]&&
	 +x_2 (8 \xt^3+12 \xo^3+8 \xo^2 \xt-8 \xo \xt^2)
	 +{1 \over 2 x_2} (12 \xo^3 \xt^2+12 \xo^2 \xt^3)
\nonumber\\[2ex]&&
	 +12 \xo^3 \xt +20 \xo^2 \xt^2+8 \xo \xt^3
          \Bigg)
+x_1^4 \Bigg(
      { 4 \over \xt} 
      \Bigg)
+x_1^3 \Bigg(
       12 -8 {\xo \over \xt}
       \Bigg)
\nonumber\\[2ex]&&
+x_1^2 \Bigg(
       24 \xo +4 {\xo^2 \over \xt}-4 \xt
       \Bigg)
+x_1 \Bigg(
       8 \xo \xt +36 \xo^2 -4 {\xo^3 \over \xt}
       +x_2 \Bigg(
            -4 {\xt^2 \over \xo} 
\nonumber\\[2ex]&&
	    +20 \xo 
	     -4 {\xo^2 \over \xt}
	     +20 \xt
	     \Bigg)
      +x_2^2 \Bigg(
            -8 {\xt \over \xo}-12 {\xo \over \xt} +40 
	    \Bigg)
      +x_2^3\Bigg(
            {4 \over \xo}-{4 \over \xt}
	    \Bigg)
      \Bigg)
\nonumber\\[2ex]&&
      +10 \xo \xt^2 
+10 \xo^2 \xt +2 \xo^3 + 2 \xt^3
\Bigg]
\Bigg\}
+ (1 \leftrightarrow 2)\,,
\nonumber\\[2ex] 
C_{qg}^{H,(1)}(\xo,\xt)&=& C_F \Bigg\{
\int_{\xo}^1~ dx_1~ {H_{qg}(x_1,\xt,\mu_F^2) \over (\xo+\xt)}
          \Bigg[
	      \Bigg(
	      {2 \xo \over x_1^2} -{4 \over x_1} + {4 \over \xo} 
	      \Bigg) {\cal L}_{a_1}
	  +2 {\xo \over x_1^2}
	  \Bigg]
\nonumber\\[2ex]&&
+\int_{\xo}^1~d x_1 ~\int_{\xt}^1~dx_2~
          {H_{qg,2}(x_1,x_2,\mu_F^2) \over (x_2-\xt) (\xo + \xt)}
	  \Bigg[
	  {2 \xo \over x_1^2} -{4 \over x_1} + {4 \over \xo} 
	  \Bigg]
\nonumber\\[2ex]&&
+\int_{\xo}^1~dx_1 ~\int_{\xt}^1~dx_2
         {H_{qg}(x_1,x_2,\mu_F^2)}
\Bigg[
	 {1 \over (\xo+\xt) (x_2+\xo)}
	   \Bigg({2 \xt \over x_1^2}+{4 \over x_1}+{4 \over \xt}
	   \Bigg)
\nonumber\\[2ex]&&
        -{2 \over x_1^2 x_2} +{4 \over x_1 x_2^2}
	+{1 \over (x_1+x_2)^2} 
	   \Bigg(
	  -{x_1 \over \xo \xt} +{x_2 \over \xo \xt} -{2 \over x_2}
	  +{2 \over \xo \xt}(\xo-\xt)
	  \Bigg)
\nonumber\\[2ex]&&
        +{1 \over (x_1+x_2)} 
	    \Bigg(
	      -{4 \over x_2^2} -{3 \over \xo \xt}
            \Bigg)
\Bigg]
\Bigg\}\,,
\nonumber\\[2ex]
C_{gq}^{H,(1)}(\xo,\xt)&=& 
C_{qg}^{H,(1)}(\xo,\xt)\Bigg|_{1 \leftrightarrow 2}\,,
\nonumber\\[2ex]
C_{q\overline q}^{H,(1)}(\xo,\xt)&=& C_F^2 \Bigg\{ 
           \int_{\xo}^1~dx_1~\int_{\xt}^1~dx_2~
	   {H_{q\overline q}(x_1,x_2,\mu_F^2) \over (\xo+\xt) }
\Bigg[
           {1 \over (x_1+x_2)^3} 
	   \Bigg(
	   {1 \over x_1^2}(4 \xo \xt^2 +4 \xo^2 \xt)
\nonumber\\[2ex]&&
	   +{1 \over x_1} (8 \xo^2-8 \xt^2)
	   +x_1 x_2 \Bigg(-{2 \over \xo}-{2 \over \xt}\Bigg)
	   +x_1 \Bigg(8 {\xt \over \xo}-8{\xo \over \xt}\Bigg)
	   +x_1^2 \Bigg({2 \over \xo}+{2 \over \xt}\Bigg)
\nonumber\\[2ex]&&
	   +4 {\xt^2 \over \xo}-12 \xo +4{\xo^2 \over \xt}-12 \xt
	   \Bigg)
        +{1 \over (x_1+x_2)} \Bigg({1 \over \xo}+{1 \over \xt}\Bigg) 
\Bigg]\Bigg\}
+ (1 \leftrightarrow 2)\,.
\end{eqnarray}

We have introduced the following abbreviations
\begin{eqnarray}
H_{ab,12}(x_1,x_2,\mu_F^2)&=&H_{ab}(x_1,x_2,\mu_F^2)
-H_{ab}(\xo,x_2,\mu_F^2)-H_{ab}(x_1,\xt,\mu_F^2)
\nonumber\\[2ex]&&
+H_{ab}(\xo,\xt,\mu_F^2)\,,
\nonumber\\[2ex]
H_{ab,1}(x_1,z,\mu_F^2)&=&H_{ab}(x_1,z,\mu_F^2)
-H_{ab}(\xo,z,\mu_F^2)\,,
\nonumber\\[2ex]
H_{ab,2}(z,x_2,\mu_F^2)&=&H_{ab}(z,x_2,\mu_F^2)
-H_{ab}(z,\xt,\mu_F^2)\,,
\label{eqa5}
\end{eqnarray}
\begin{eqnarray}
{\cal L}_{a_1}&=&\ln\left(
{q^2 (\xo+\xt) (1-\xt)(x_1-\xo) \over
\mu_F^2 (x_1+\xt) \xo \xt}
\right)\,,
\quad \quad
{\cal L}_{b_1}=\ln\left(
{q^2 (1-\xt)(x_1-\xo) \over
\mu_F^2  \xo \xt}
\right)\,,
\nonumber\\[2ex]
{\cal L}_{c_1}&=&\ln\left(
{ \xo+\xt \over
x_1+\xt}
\right)\,,
\quad \quad
{\cal L}(\xo,\xt)=\ln\left(
{ (1-\xo) (1-\xt) \over
\xo \xt}
\right)\,.
\label{eqa6}
\end{eqnarray}
The $D_{ab}^{H,(i)}(\xo,\xt)$ that appear in the hard parts of the 
rapidity distributions are listed below
\begin{eqnarray}
D_{gg}^{H,(0)}(\xo,\xt)&=& H_{gg}(\xo,\xt,\mu_F^2)\,,
\nonumber\\[2ex]
D_{gg}^{H,(1)}(\xo,\xt)&=& C_A \Bigg\{
        \int_{\xo}^1~dx_1~ H_{gg}(x_1,\xt,\mu_F^2) 
	\Bigg[\Bigg(
	  -4 {\xo^2 \over x_1^3}+4 {\xo \over x_1^2}-{8 \over x_1}+{4 \over \xo}
        \Bigg){\cal K}_{a_1}
\nonumber\\[2ex]&&
	+{4 \over (x_1 -\xo)} {\cal K}_{c_1}
	\Bigg]
+\int_{\xo}^1~dx_1 ~
        {H_{gg,1}(x_1,\xt,\mu_F^2) \over (x_1-\xo)}
	\Bigg[4 {\cal K}_{b_1}\Bigg]
\nonumber\\[2ex]&&
+\int_{\xo}^1~dx_1~\int_{\xt}^1~dx_2~
        {H_{gg,1}(x_1,x_2,\mu_F^2)\over (x_1-\xo)}
	\Bigg[
	-4 {\xt^2 \over x_2^3}+4 {\xt \over x_2^2}-{8 \over x_2}+{4 \over \xt}
	\Bigg]
\nonumber\\[2ex]&&
+\int_{\xo}^1~dx_1~\int_{\xt}^1~dx_2~
         {2 H_{gg,12}(x_1,x_2,\mu_F^2) \over (x_1 -\xo ) (x_2 -\xt)}
\nonumber\\[2ex]&&
+H_{gg}(\xo,\xt,\mu_F^2)
          \Bigg[
	  2 {\cal K}(\xo,\xt) \log\left({q^2 \over \mu_F^2}\right) 
	   +6 \zeta_2
	   +\left({\cal K}(\xo,\xt)\right)^2
	  \Bigg]
\nonumber\\[2ex]&&
+\int_{\xo}^1~dx_1~\int_{\xt}^1~dx_2~
     {H_{gg}(x_1,x_2,\mu_F^2) \over (x_1+\xo)(x_2+\xt)(x_1 \xt+ \xo x_2)^4 }
     \Bigg[
       {1 \over x_1^3} 
           \Bigg(
	   4 x_2^4 \xo^7 +8 x_2^3 \xo^7 \xt 
\nonumber\\[2ex]&&
	   +8 x_2^2 \xo^7 \xt^2 +8 x_2 \xo^7 \xt^3
           \Bigg)
      + {1 \over x_1^2} 
           \Bigg(
           16 \xo^6 \xt^4+16 x_2^3 \xo^6 \xt + 32 x_2^2 \xo^6 \xt^2 
\nonumber\\[2ex]&&
	   +24 x_2 \xo^6 \xt^3
           \Bigg)
      + {1 \over x_1} 
           \Bigg(
	   40 \xo^5 \xt^4+4 x_2^4 \xo^5+ 8 x_2^3 \xo^5 \xt+32 x_2^2 \xo^5 \xt^2
\nonumber\\[2ex]&&
	        +48 x_2 \xo^5 \xt^3 +{12 \over x_2} \xo^5 \xt^5
           \Bigg)
      + x_1^5 
           \Bigg(
	   4 {\xt^4 \over \xo}
           \Bigg)
      + x_1^4 
           \Bigg(
	   4 \xt^4
           \Bigg)
      + x_1^3 
           \Bigg(
	   12 x_2^3 \xo \xt 
\nonumber\\[2ex]&&
	   +28 \xo \xt^4
           \Bigg)
      + x_1^2 
           \Bigg(
	   48 \xo^2 \xt^4+16 x_2^4 \xo^2 +20 x_2^2 \xo^2 \xt^2+88 x_2 \xo^2 \xt^3
           \Bigg)
\nonumber\\[2ex]&&
      + x_1 
           \Bigg(
	   68 \xo^3 \xt^4+{8 x_2^5 \xo^3 \over \xt}+ 4 x_2^4 \xo^3 +8 x_2^3 \xo^3 \xt
	      +64 x_2 \xo^3 \xt^3
           \Bigg)
\nonumber\\[2ex]&&
     +40 \xo^4 \xt^4
     \Bigg]
\Bigg\}
+ (1 \leftrightarrow 2) \,,
\nonumber\\[2ex]
D_{qg}^{H,(1)}(\xo,\xt)&=& C_F \Bigg\{
\int_{\xo}^1~dx_1~ H_{qg}(x_1,\xt,\mu_F^2) 
      \Bigg[
       \Bigg(
      {2 \xo \over x_1^2}-{4 \over x_1}+{4 \over \xo}
      \Bigg){\cal K}_{a_1}
      +{2 \xo \over x_1^2} 
      \Bigg]
\nonumber\\[2ex]&&
   + \int_{\xo}^1~dx_1~\int_{\xt}^1~dx_2~
     {H_{qg,2}(x_1,x_2,\mu_F^2) \over (x_2-\xt)}
     \Bigg[
     {2 \xo \over x_1^2}-{4 \over x_1} + {4 \over \xo}
     \Bigg]
\nonumber\\[2ex]&&
+\int_{\xo}^1~dx_1~\int_{\xt}^1~dx_2~{H_{qg}(x_1,x_2,\mu_F^2) \over (x_1+\xo) (x_2+\xt) (x_1 \xt+\xo x_2)^3}
      \Bigg[
    {1 \over x_1^2} 
        \Bigg(
	-2 x_2^3 \xo^5 
\nonumber\\[2ex]&&
	-4 x_2^2 \xo^5 \xt -4 x_2 \xo^5 \xt^2
	\Bigg)
    +{1 \over x_1} 
        \Bigg(
	2 x_2^3 \xo^4-2 x_2^2 \xo^4 \xt-4 x_2 \xo^4 \xt^2
	\Bigg)
\nonumber\\[2ex]&&
    + x_1 
        \Bigg(
	6 x_2 \xo^2 \xt^2+2 \xo^2 \xt^3
	\Bigg)
    + x_1^2 
        \Bigg(
	-12 x_2 \xo \xt^2 +4 x_2^3 \xo - 2 \xo \xt^3
	\Bigg)
\nonumber\\[2ex]&&
    + x_1^3 
        \Bigg(
	4 x_2 \xt^2+8 x_2^2 \xt
	\Bigg)
    + x_1^4 
        \Bigg(
	8 {x_2 \xt^2 \over \xo}+4 {\xt^3 \over \xo} 
	\Bigg)
    +14 x_2 \xo^3 \xt^2 
\nonumber\\[2ex]&&
    +10 x_2^2 \xo^3 \xt
      \Bigg]
\Bigg\}\,,
\nonumber\\[2ex]
D_{gq}^{H,(1)}(\xo,\xt)&=& 
D_{qg}^{H,(1)}(\xo,\xt)\Bigg|_{1\leftrightarrow 2}\,,
\nonumber\\[2ex]
D_{q\overline q}^{H,(1)}(\xo,\xt)&=& C_F^2 \Bigg\{
\int_{\xo}^1~ dx_1 ~\int_{\xt}^1~dx_2~ 
        {H_{q \overline q}(x_1,x_2,\mu_F^2) \over (x_1+\xo)(x_2+\xt)(x_1 \xt + \xo x_2)^4}
\Bigg[
    {1 \over x_1^2}
       \Bigg(
       8 x_2 \xo^6 \xt^3 
\nonumber\\[2ex]&&
       + 8 \xo^6 \xt^4
       \Bigg)
   + {1 \over x_1}
       \Bigg(
       16 x_2 \xo^5 \xt^3+8 x_2^2 \xo^5 \xt^2+8 \xo^5 \xt^4
       \Bigg)
   + x_1
       \Bigg(
       -32 x_2 \xo^3 \xt^3 
\nonumber\\[2ex]&&
       -24 x_2^2 \xo^3 \xt^2+16 x_2^3 \xo^3 \xt
          +8 x_2^4 \xo^3 -24 \xo^3 \xt^4
       \Bigg)
   + x_1^2
       \Bigg(
       -16 x_2^2 \xo^2 \xt^2 
\nonumber\\[2ex]&&
       +8 x_2^3 \xo^2 \xt +8 x_2^4 \xo^2+16 \xo^2 \xt^4
       \Bigg)
   + x_1^3
       \Bigg(
       8 \xo \xt^4
       \Bigg)
   -16 \xo^4 \xt^4
\Bigg]
\Bigg\}
\nonumber\\[2ex]&&
+ (1 \leftrightarrow 2)\,,
\end{eqnarray}
where we have introduced the following abbreviations
\begin{eqnarray}
{\cal K}_{a_1}&=&\ln\left(
{2 q^2 (1-\xt)(x_1-\xo) \over
\mu_F^2 (x_1+\xo)  \xt}
\right)\,,
\quad \quad
{\cal K}_{b_1}=\ln\left(
{q^2 (1-\xt)(x_1-\xo) \over
\mu_F^2  \xo \xt}
\right)\,,
\nonumber\\[2ex]
{\cal K}_{c_1}&=&\ln\left(
{ 2 \xo \over
x_1+\xo}
\right)\,,
\quad \quad
{\cal K}(\xo,\xt)=\ln\left(
{ (1-\xo) (1-\xt) \over
\xo \xt}
\right)\,.
\label{eqb5}
\end{eqnarray}
The ${\cal K}_{a_2}$, ${\cal K}_{b_2}$ and ${\cal K}_{c_2}$ can be obtained
from ${\cal K}_{a_1}$, ${\cal K}_{b_1}$ and ${\cal K}_{c_1}$
by using $1 \leftrightarrow 2$ symmetry.

\section{Soft-plus-virtual parts}
Here we list below the $\Delta_{Y,a}^{{\rm sv},(i)}$ that contribute 
to the soft-plus-virtual parts
of the cross sections for the choice $\mu_R^2=\mu_F^2=q^2$.
\begin{eqnarray}
  \Delta^{{\rm sv},(1)}_{Y,q} &=&
        \delta(1-z_1) \delta(1-z_2) \Bigg[C_F   \Bigg(
          - 8
          + 6 ~\zeta_2
          \Bigg)
\Bigg]
       + {\cal D}_0 {\cal \overline D}_0 \Bigg[
          C_F   \Bigg(
           2
          \Bigg)
\Bigg]
       + {\cal \overline D}_1 \delta(1-z_1) \Bigg[
          C_F   \Bigg(
           4
          \Bigg)
\Bigg]
\nonumber\\&& + (z_1 \leftrightarrow z_2) \,,
\nonumber\\
   \Delta^{{\rm sv},(2)}_{Y,q} &=&
\delta(1-z_1) \delta(1-z_2) \Bigg[ n_f C_F   \Bigg(
           {127 \over 12}
          - {76 \over 9} ~\zeta_2
          + {4 \over 3} ~\zeta_3
          \Bigg)
       +  C_F C_A   \Bigg(
          - {1535 \over 24}
\nonumber\\&&
          + {430 \over 9} ~\zeta_2
          - {26 \over 5} ~\zeta_2^2
          + {86 \over 3} ~\zeta_3
          \Bigg)
       +  C_F^2   \Bigg(
          { 511 \over 8}
          - 67 ~\zeta_2
          + {152 \over 5} ~\zeta_2^2
          - 30 ~\zeta_3
          \Bigg)
\Bigg]
\nonumber\\&&
+ {\cal D}_0 {\cal \overline D}_0 \Bigg[
         n_f C_F   \Bigg(
          - {20 \over 9}
          \Bigg)
       +  C_F C_A   \Bigg(
           {134 \over 9}
          - 4 ~\zeta_2
          \Bigg)
       +  C_F^2   \Bigg(
          - 32
          + 8 ~\zeta_2
          \Bigg)
\Bigg]
\nonumber\\&&
+ {\cal D}_0 {\cal \overline D}_1 \Bigg[
         n_f C_F   \Bigg(
           {8 \over 3}
          \Bigg)
       + C_F C_A   \Bigg(
          - {44 \over 3}
          \Bigg)
\Bigg]
+ {\cal D}_0 {\cal \overline D}_2 \Bigg[ 
        C_F^2   \Bigg(
           24
          \Bigg)
\Bigg]
+ {\cal D}_1 {\cal \overline D}_1 \Bigg[
          C_F^2   \Bigg(
           24
          \Bigg)
\Bigg]
\nonumber\\&&
+ {\cal \overline D}_0 \delta(1-z_1) \Bigg[
          n_f C_F   \Bigg(
           {112 \over 27}
          - {8 \over 3} ~\zeta_2
          \Bigg)
       +  C_F C_A   \Bigg(
          - {808 \over 27}
          + {44 \over 3} ~\zeta_2
          + 28 ~\zeta_3
          \Bigg)
\nonumber\\&&
       +  C_F^2   \Bigg(
           32 ~\zeta_3
          \Bigg)
\Bigg]
+ {\cal \overline D}_1 \delta(1-z_1) \Bigg[
	  n_f C_F   \Bigg(
          - {40 \over 9}
          \Bigg)
       +  C_F C_A   \Bigg(
           {268 \over 9}
          - 8 ~\zeta_2
          \Bigg)
\nonumber\\&&
       +  C_F^2   \Bigg(
          - 64
          + 16 ~\zeta_2
          \Bigg)
\Bigg]
       + {\cal \overline D}_2 \delta(1-z_1) \Bigg[
          n_f C_F   \Bigg(
           {4 \over 3}
          \Bigg)
       +  C_F C_A   \Bigg(
          - {22 \over 3}
          \Bigg)
\Bigg]
\nonumber\\&&
       + {\cal \overline D}_3 \delta(1-z_1)\Bigg[
          C_F^2   \Bigg(
           8
          \Bigg)
\Bigg]
\nonumber\\&& + (z_1 \leftrightarrow z_2) \,,
\nonumber\\[2ex]
   \Delta^{{\rm sv},(3)}_{Y,q} &=&
{\cal D}_0 {\cal \overline D}_0  \Bigg[
         n_f C_F C_A   \Bigg(
          - {4102 \over 81}
          + {256 \over 9} ~\zeta_2
          \Bigg)
       +  n_f C_F^2   \Bigg(
           {536 \over 9}
          - {224 \over 9} ~\zeta_2
          + {160 \over 3} ~\zeta_3
          \Bigg)
\nonumber\\&&
       + n_f^2 C_F   \Bigg(
           {200 \over 81}
          - {16 \over 9} ~\zeta_2
          \Bigg)
       + C_F C_A^2   \Bigg(
           {15503 \over 81}
          - {340 \over 3} ~\zeta_2
          + {88 \over 5} ~\zeta_2^2
          - 88 ~\zeta_3
          \Bigg)
\nonumber\\&&
       +  C_F^2 C_A   \Bigg(
          - {8893 \over 18}
          + {1760 \over 9} ~\zeta_2
          - {24 \over 5} ~\zeta_2^2
          - {184 \over 3} ~\zeta_3
          \Bigg)
       +  C_F^3   \Bigg(
           {511 \over 2}
          - 12 ~\zeta_2
\nonumber\\&&
          - {96 \over 5} ~\zeta_2^2
          - 120 ~\zeta_3
          \Bigg)
\Bigg]
+ {\cal D}_0 {\cal \overline D}_1 \Bigg[
          n_f C_F C_A   \Bigg(
           {2312 \over 27}
          - {32 \over 3} ~\zeta_2
          \Bigg)
       +  n_f C_F^2   \Bigg(
           {136 \over 9}
\nonumber\\&&
          - 64 ~\zeta_2
          \Bigg)
       +  n_f^2 C_F   \Bigg(
          - {160 \over 27}
          \Bigg)
       +  C_F C_A^2   \Bigg(
          - {7120 \over 27}
          + {176 \over 3} ~\zeta_2
          \Bigg)
       +  C_F^2 C_A   \Bigg(
          - {1120 \over 9}
\nonumber\\&&
          + 352 ~\zeta_2
          + 336 ~\zeta_3
          \Bigg)
       +  C_F^3   \Bigg(
           640 ~\zeta_3
          \Bigg)
\Bigg]
+ {\cal D}_0 {\cal \overline D}_2 \Bigg[
          n_f C_F C_A   \Bigg(
          - {176 \over 9}
          \Bigg)
       +  n_f C_F^2   \Bigg(
          - {160 \over 3}
          \Bigg)
\nonumber\\&&
       +  n_f^2 C_F   \Bigg(
           {16 \over 9}
          \Bigg)
       +  C_F C_A^2   \Bigg(
           {484 \over 9}
          \Bigg)
       +  C_F^2 C_A   \Bigg(
           {1072 \over 3}
          - 96 ~\zeta_2
          \Bigg)
       +  C_F^3   \Bigg(
          - 384
          - 96 ~\zeta_2
          \Bigg)
\Bigg]
\nonumber\\&&
+ {\cal D}_0 {\cal \overline D}_3 \Bigg[
          n_f C_F^2   \Bigg(
           {160 \over 9}
          \Bigg)
       +  C_F^2 C_A   \Bigg(
          - {880 \over 9}
          \Bigg)
\Bigg]
       + {\cal D}_0 {\cal \overline D}_4 \Bigg[
          C_F^3   \Bigg(
           40
          \Bigg)
\Bigg]
\nonumber\\&&
+ {\cal D}_1 {\cal \overline D}_1 \Bigg[
         n_f C_F C_A   \Bigg(
          - {176 \over 9}
          \Bigg)
       +  n_f C_F^2   \Bigg(
          - {160 \over 3}
          \Bigg)
       +  n_f^2 C_F   \Bigg(
           {16 \over 9}
          \Bigg)
       +  C_F C_A^2   \Bigg(
           {484 \over 9}
          \Bigg)
\nonumber\\&&
       +  C_F^2 C_A   \Bigg(
           {1072 \over 3}
          - 96 ~\zeta_2
          \Bigg)
       +  C_F^3   \Bigg(
          - 384
          - 96 ~\zeta_2
          \Bigg)
\Bigg]
+ {\cal D}_1 {\cal \overline D}_2 \Bigg[
         n_f C_F^2   \Bigg(
           {160 \over 3}
          \Bigg)
\nonumber\\&&
       +  C_F^2 C_A   \Bigg(
          - {880 \over 3}
          \Bigg)
\Bigg]
+ {\cal D}_1 {\cal \overline D}_3 \Bigg[
          C_F^3   \Bigg(
           160
          \Bigg)
\Bigg]
+ {\cal D}_2 {\cal \overline D}_2 \Bigg[
          C_F^3   \Bigg(
           120
          \Bigg)
\Bigg]
\nonumber\\&&
+ {\cal \overline D}_0 \delta(1-z_1) \Bigg[
           n_f C_F C_A   \Bigg(
           {62626 \over 729}
          - {7760 \over 81} ~\zeta_2
          + {208 \over 15} ~\zeta_2^2
          - {536 \over 9} ~\zeta_3
          \Bigg)
\nonumber\\&&
       +  n_f C_F^2   \Bigg(
          - 3
          + {1384 \over 27} ~\zeta_2
          - {256 \over 15} ~\zeta_2^2
          - {944 \over 9} ~\zeta_3
          \Bigg)
       +  n_f^2 C_F   \Bigg(
          - {1856 \over 729}
          + {160 \over 27} ~\zeta_2
\nonumber\\&&
          - {32 \over 27} ~\zeta_3
          \Bigg)
       +  C_F C_A^2   \Bigg(
          - {297029 \over 729}
          - {176 \over 3} ~\zeta_2 ~\zeta_3
          + {27752 \over 81} ~\zeta_2
          - {616 \over 15} ~\zeta_2^2
\nonumber\\&&
          + {14264 \over 27} ~\zeta_3
          - 192 ~\zeta_5
          \Bigg)
       +  C_F^2 C_A   \Bigg(
           {12928 \over 27}
          - 16 ~\zeta_2 ~\zeta_3
          - {9568 \over 27} ~\zeta_2
          + {176 \over 3} ~\zeta_2^2
\nonumber\\&&
          + {256 \over 9} ~\zeta_3
          \Bigg)
       +  C_F^3   \Bigg(
          - 128 ~\zeta_2 ~\zeta_3
          - 512 ~\zeta_3
          + 384 ~\zeta_5
          \Bigg)
\Bigg]
\nonumber\\&&
+ {\cal \overline D}_1 \delta(1-z_1) \Bigg[ 
           n_f C_F C_A   \Bigg(
          - {8204 \over 81}
          + {512 \over 9} ~\zeta_2
          \Bigg)
       +  n_f C_F^2   \Bigg(
           {1072 \over 9}
          - {448 \over 9} ~\zeta_2
\nonumber\\&&
          + {320 \over 3} ~\zeta_3
          \Bigg)
       +  n_f^2 C_F   \Bigg(
           {400 \over 81}
          - {32 \over 9} ~\zeta_2
          \Bigg)
       +  C_F C_A^2   \Bigg(
           {31006 \over 81}
          - {680 \over 3} ~\zeta_2
          + {176 \over 5} ~\zeta_2^2
\nonumber\\&&
          - 176 ~\zeta_3
          \Bigg)
       +  C_F^2 C_A   \Bigg(
          - {8893 \over 9}
          + {3520 \over 9} ~\zeta_2
          - {48 \over 5} ~\zeta_2^2
          - {368 \over 3} ~\zeta_3
          \Bigg)
\nonumber\\&&
       +  C_F^3   \Bigg(
           511
          - 24 ~\zeta_2
          - {192 \over 5} ~\zeta_2^2
          - 240 ~\zeta_3
          \Bigg)
\Bigg]
+ {\cal \overline D}_2 \delta(1-z_1) \Bigg[
           n_f C_F C_A   \Bigg(
           {1156 \over 27}
\nonumber\\&&
          - {16 \over 3} ~\zeta_2
          \Bigg)
       +  n_f C_F^2   \Bigg(
           {68 \over 9}
          - 32 ~\zeta_2
          \Bigg)
       +  n_f^2 C_F   \Bigg(
          - {80 \over 27}
          \Bigg)
       +  C_F C_A^2   \Bigg(
          - {3560 \over 27}
          + {88 \over 3} ~\zeta_2
          \Bigg)
\nonumber\\&&
       +  C_F^2 C_A   \Bigg(
          - {560 \over 9}
          + 176 ~\zeta_2
          + 168 ~\zeta_3
          \Bigg)
       +  C_F^3   \Bigg(
           320 ~\zeta_3
          \Bigg)
\Bigg]
\nonumber\\&&
+ {\cal \overline D}_3 \delta(1-z_1) \Bigg[
          n_f C_F C_A   \Bigg(
          - {176 \over 27}
          \Bigg)
       +  n_f C_F^2   \Bigg(
          - {160 \over 9}
          \Bigg)
       +  n_f^2 C_F   \Bigg(
           {16 \over 27}
          \Bigg)
\nonumber\\&&
       +  C_F C_A^2   \Bigg(
           {484 \over 27}
          \Bigg)
       +  C_F^2 C_A   \Bigg(
           {1072 \over 9}
          - 32 ~\zeta_2
          \Bigg)
       +  C_F^3   \Bigg(
          - 128
          - 32 ~\zeta_2
          \Bigg)
\Bigg]
\nonumber\\&&
+ {\cal \overline D}_4 \delta(1-z_1)\Bigg[ n_f C_F^2   \Bigg(
           {40 \over 9}
          \Bigg)
       +  C_F^2 C_A   \Bigg(
          - {220 \over 9}
          \Bigg)
\Bigg]
+ {\cal \overline D}_5 \delta(1-z_1) \Bigg[
          C_F^3   \Bigg(
           8
          \Bigg)
\Bigg]
\nonumber\\&& + (z_1 \leftrightarrow z_2) \,,
\nonumber\\[2ex]
   \Delta^{{\rm sv},(4)}_{Y,q} &=&
 {\cal D}_0 {\cal \overline D}_1 \Bigg[
          n_f C_F C_A^2   \Bigg(
           {20554 \over 9}
          - {10192 \over 9} ~\zeta_2
          + {352 \over 5} ~\zeta_2^2
          - 352 ~\zeta_3
          \Bigg)
       +  n_f C_F^2 C_A   \Bigg(
           {84280 \over 243}
\nonumber\\&&
          - {26432 \over 9} ~\zeta_2
          + {1376 \over 3} ~\zeta_2^2
          - {27424 \over 9} ~\zeta_3
          \Bigg)
       +  n_f C_F^3   \Bigg(
           {518 \over 3}
          + {7088 \over 9} ~\zeta_2
          - {2176 \over 15} ~\zeta_2^2
\nonumber\\&&
          - {8096 \over 3} ~\zeta_3
          \Bigg)
       +  n_f^2 C_F C_A   \Bigg(
          - {7894 \over 27}
          + {1376 \over 9} ~\zeta_2
          \Bigg)
       +  n_f^2 C_F^2   \Bigg(
          - {7436 \over 243}
          + {5504 \over 27} ~\zeta_2
\nonumber\\&&
          + {640 \over 3} ~\zeta_3
          \Bigg)
       +  n_f^3 C_F   \Bigg(
           {800 \over 81}
          - {64 \over 9} ~\zeta_2
          \Bigg)
       +  C_F C_A^3   \Bigg(
          - {412880 \over 81}
          + {8024 \over 3} ~\zeta_2
\nonumber\\&&
          - {1936 \over 5} ~\zeta_2^2
          + 1936 ~\zeta_3
          \Bigg)
       +  C_F^2 C_A^2   \Bigg(
          - {356573 \over 243}
          - 1376 ~\zeta_2 ~\zeta_3
          + {258784 \over 27} ~\zeta_2
\nonumber\\&&
          - {31504 \over 15} ~\zeta_2^2
          + {114592 \over 9} ~\zeta_3
          - 2304 ~\zeta_5
          \Bigg)
       +  C_F^3 C_A   \Bigg(
           {34849 \over 9}
          - 5184 ~\zeta_2 ~\zeta_3
\nonumber\\&&
          - {45416 \over 9} ~\zeta_2
          + {5632 \over 15} ~\zeta_2^2
          + {29392 \over 3} ~\zeta_3
          \Bigg)
       +  C_F^4   \Bigg(
          - 7680 ~\zeta_2 ~\zeta_3
          - 10240 ~\zeta_3
\nonumber\\&&
          + 10752 ~\zeta_5
          \Bigg)
\Bigg]
+ {\cal D}_0 {\cal \overline D}_2 \Bigg[
         n_f C_F C_A^2   \Bigg(
          - {7324 \over 9}
          + {352 \over 3} ~\zeta_2
          \Bigg)
       +  n_f C_F^2 C_A   \Bigg(
          - {133988 \over 81}
\nonumber\\&&
          + {14144 \over 9} ~\zeta_2
          + {448 \over 3} ~\zeta_3
          \Bigg)
       +  n_f C_F^3   \Bigg(
           {2764 \over 3}
          + {704 \over 3} ~\zeta_2
          + {5696 \over 3} ~\zeta_3
          \Bigg)
\nonumber\\&&
       +  n_f^2 C_F C_A   \Bigg(
           {1144 \over 9}
          - {32 \over 3} ~\zeta_2
          \Bigg)
       +  n_f^2 C_F^2   \Bigg(
           {7768 \over 81}
          - {1024 \over 9} ~\zeta_2
          \Bigg)
       +  n_f^3 C_F   \Bigg(
          - {160 \over 27}
          \Bigg)
\nonumber\\&&
       +  C_F C_A^3   \Bigg(
           {43648 \over 27}
          - {968 \over 3} ~\zeta_2
          \Bigg)
       +  C_F^2 C_A^2   \Bigg(
           {481216 \over 81}
          - {50272 \over 9} ~\zeta_2
          + {2592 \over 5} ~\zeta_2^2
\nonumber\\&&
          - {8800 \over 3} ~\zeta_3
          \Bigg)
       +  C_F^3 C_A   \Bigg(
          - {26362 \over 3}
          - {1376 \over 3} ~\zeta_2
          + {4512 \over 5} ~\zeta_2^2
          - {19808 \over 3} ~\zeta_3
          \Bigg)
\nonumber\\&&
       +  C_F^4   \Bigg(
           3066
          + 2928 ~\zeta_2
          - {4992 \over 5} ~\zeta_2^2
          - 1440 ~\zeta_3
          \Bigg)
\Bigg]
+ {\cal D}_0 {\cal \overline D}_3 \Bigg[
         n_f C_F C_A^2   \Bigg(
           {968 \over 9}
          \Bigg)
\nonumber\\&&
       +  n_f C_F^2 C_A   \Bigg(
           {21920 \over 27}
          - {320 \over 3} ~\zeta_2
          \Bigg)
       +  n_f C_F^3   \Bigg(
          - {1760 \over 27}
          - {4160 \over 9} ~\zeta_2
          \Bigg)
\nonumber\\&&
       +  n_f^2 C_F C_A   \Bigg(
          - {176 \over 9}
          \Bigg)
       +  n_f^2 C_F^2   \Bigg(
          - {1600 \over 27}
          \Bigg)
       +  n_f^3 C_F   \Bigg(
           {32 \over 27}
          \Bigg)
       +  C_F C_A^3   \Bigg(
          - {5324 \over 27}
          \Bigg)
\nonumber\\&&
       +  C_F^2 C_A^2   \Bigg(
          - {67120 \over 27}
          + {1760 \over 3} ~\zeta_2
          \Bigg)
       +  C_F^3 C_A   \Bigg(
           {9920 \over 27}
          + {22880 \over 9} ~\zeta_2
          + 1120 ~\zeta_3
          \Bigg)
\nonumber\\&&
       +  C_F^4   \Bigg(
           {8960 \over 3} ~\zeta_3
          \Bigg)
\Bigg]
+ {\cal D}_0 {\cal \overline D}_4 \Bigg[
         n_f C_F^2 C_A   \Bigg(
          - {3520 \over 27}
          \Bigg)
       +  n_f C_F^3   \Bigg(
          - {400 \over 3}
          \Bigg)
\nonumber\\&&
       +  n_f^2 C_F^2   \Bigg(
           {320 \over 27}
          \Bigg)
       +  C_F^2 C_A^2   \Bigg(
           {9680 \over 27}
          \Bigg)
       +  C_F^3 C_A   \Bigg(
           {2680 \over 3}
          - 240 ~\zeta_2
          \Bigg)
       +  C_F^4   \Bigg(
          - 640
\nonumber\\&&
          - 480 ~\zeta_2
          \Bigg)
\Bigg]
+ {\cal D}_0 {\cal \overline D}_5 \Bigg[
         n_f C_F^3   \Bigg(
           {112 \over 3}
          \Bigg)
       +  C_F^3 C_A   \Bigg(
          - {616 \over 3}
          \Bigg)
\Bigg]
+ {\cal D}_0 {\cal \overline D}_6 \Bigg[C_F^4   \Bigg(
           {112 \over 3}
          \Bigg)
\Bigg]
\nonumber\\&&
+ {\cal D}_1 {\cal \overline D}_1 \Bigg[
           n_f C_F C_A^2   \Bigg(
          - {7324 \over 9}
          + {352 \over 3} ~\zeta_2
          \Bigg)
       +  n_f C_F^2 C_A   \Bigg(
          - {133988 \over 81}
          + {14144 \over 9} ~\zeta_2
\nonumber\\&&
          + {448 \over 3} ~\zeta_3
          \Bigg)
       +  n_f C_F^3   \Bigg(
           {2764 \over 3}
          + {704 \over 3} ~\zeta_2
          + {5696 \over 3} ~\zeta_3
          \Bigg)
       +  n_f^2 C_F C_A   \Bigg(
           {1144 \over 9}
          - {32 \over 3} ~\zeta_2
          \Bigg)
\nonumber\\&&
       +  n_f^2 C_F^2   \Bigg(
           {7768 \over 81}
          - {1024 \over 9} ~\zeta_2
          \Bigg)
       +  n_f^3 C_F   \Bigg(
          - {160 \over 27}
          \Bigg)
       +  C_F C_A^3   \Bigg(
           {43648 \over 27}
          - {968 \over 3} ~\zeta_2
          \Bigg)
\nonumber\\&&
       +  C_F^2 C_A^2   \Bigg(
           {481216 \over 81}
          - {50272 \over 9} ~\zeta_2
          + {2592 \over 5} ~\zeta_2^2
          - {8800 \over 3} ~\zeta_3
          \Bigg)
       +  C_F^3 C_A   \Bigg(
          - {26362 \over 3}
\nonumber\\&&
          - {1376 \over 3} ~\zeta_2
          + {4512 \over 5} ~\zeta_2^2
          - {19808 \over 3} ~\zeta_3
          \Bigg)
       +  C_F^4   \Bigg(
           3066
          + 2928 ~\zeta_2
          - {4992 \over 5} ~\zeta_2^2
\nonumber\\&&
          - 1440 ~\zeta_3
          \Bigg)
\Bigg]
+ {\cal D}_1 {\cal \overline D}_2 \Bigg[
         n_f C_F C_A^2   \Bigg(
           {968 \over 3}
          \Bigg)
       +  n_f C_F^2 C_A   \Bigg(
           {21920 \over 9}
          - 320 ~\zeta_2
          \Bigg)
\nonumber\\&&
       +  n_f C_F^3   \Bigg(
          - {1760 \over 9}
          - {4160 \over 3} ~\zeta_2
          \Bigg)
       +  n_f^2 C_F C_A   \Bigg(
          - {176 \over 3}
          \Bigg)
       +  n_f^2 C_F^2   \Bigg(
          - {1600 \over 9}
          \Bigg)
\nonumber\\&&
       +  n_f^3 C_F   \Bigg(
           {32 \over 9}
          \Bigg)
       +  C_F C_A^3   \Bigg(
          - {5324 \over 9}
          \Bigg)
       +  C_F^2 C_A^2   \Bigg(
          - {67120 \over 9}
          + 1760 ~\zeta_2
          \Bigg)
\nonumber\\&&
       +  C_F^3 C_A   \Bigg(
           {9920 \over 9}
          + {22880 \over 3} ~\zeta_2
          + 3360 ~\zeta_3
          \Bigg)
       +  C_F^4   \Bigg(
           8960 ~\zeta_3
          \Bigg)
\Bigg]
\nonumber\\&&
+ {\cal D}_1 {\cal \overline D}_3 \Bigg[
         n_f C_F^2 C_A   \Bigg(
          - {14080 \over 27}
          \Bigg)
       +  n_f C_F^3   \Bigg(
          - {1600 \over 3}
          \Bigg)
       +  n_f^2 C_F^2   \Bigg(
           {1280 \over 27}
          \Bigg)
\nonumber\\&&
       +  C_F^2 C_A^2   \Bigg(
           {38720 \over 27}
          \Bigg)
       +  C_F^3 C_A   \Bigg(
           {10720 \over 3}
          - 960 ~\zeta_2
          \Bigg)
       +  C_F^4   \Bigg(
          - 2560
          - 1920 ~\zeta_2
          \Bigg)
\Bigg]
\nonumber\\&&
+ {\cal D}_1 {\cal \overline D}_4 \Bigg[
         n_f C_F^3   \Bigg(
           {560 \over 3}
          \Bigg)
       +  C_F^3 C_A   \Bigg(
          - {3080 \over 3}
          \Bigg)
\Bigg]
+ {\cal D}_1 {\cal \overline D}_5 \Bigg[ C_F^4   \Bigg(
           224
          \Bigg)
\Bigg]
\nonumber\\&&
+ {\cal D}_2 {\cal \overline D}_2 \Bigg[
         n_f C_F^2 C_A   \Bigg(
          - {3520 \over 9}
          \Bigg)
       +  n_f C_F^3   \Bigg(
          - 400
          \Bigg)
       +  n_f^2 C_F^2   \Bigg(
           {320 \over 9}
          \Bigg)
\nonumber\\&&
       +  C_F^2 C_A^2   \Bigg(
           {9680 \over 9}
          \Bigg)
       +  C_F^3 C_A   \Bigg(
           2680
          - 720 ~\zeta_2
          \Bigg)
       +  C_F^4   \Bigg(
          - 1920
          - 1440 ~\zeta_2
          \Bigg)
\Bigg]
\nonumber\\&&
+ {\cal D}_2 {\cal \overline D}_3 \Bigg[
         n_f C_F^3   \Bigg(
           {1120 \over 3}
          \Bigg)
       +  C_F^3 C_A   \Bigg(
          - {6160 \over 3}
          \Bigg)
\Bigg]
+ {\cal D}_2 {\cal \overline D}_4 \Bigg[
          C_F^4   \Bigg(
           560
          \Bigg)
\Bigg]
\nonumber\\&&
+ {\cal D}_3 {\cal \overline D}_3 \Bigg[
          C_F^4   \Bigg(
           {1120 \over 3}
          \Bigg)
\Bigg]
+ {\cal \overline D}_2 \delta(1-z_1) \Bigg[
            n_f C_F C_A^2   \Bigg(
           {10277 \over 9}
          - {5096 \over 9} ~\zeta_2
          + {176 \over 5} ~\zeta_2^2
\nonumber\\&&
          - 176 ~\zeta_3
          \Bigg)
       +  n_f C_F^2 C_A   \Bigg(
           {42140 \over 243}
          - {13216 \over 9} ~\zeta_2
          + {688 \over 3} ~\zeta_2^2
          - {13712 \over 9} ~\zeta_3
          \Bigg)
\nonumber\\&&
       +  n_f C_F^3   \Bigg(
           {259 \over 3}
          + {3544 \over 9} ~\zeta_2
          - {1088 \over 15} ~\zeta_2^2
          - {4048 \over 3} ~\zeta_3
          \Bigg)
       +  n_f^2 C_F C_A   \Bigg(
          - {3947 \over 27}
\nonumber\\&&
          + {688 \over 9} ~\zeta_2
          \Bigg)
       +  n_f^2 C_F^2   \Bigg(
          - {3718 \over 243}
          + {2752 \over 27} ~\zeta_2
          + {320 \over 3} ~\zeta_3
          \Bigg)
       +  n_f^3 C_F   \Bigg(
           {400 \over 81}
\nonumber\\&&
          - {32 \over 9} ~\zeta_2
          \Bigg)
       +  C_F C_A^3   \Bigg(
          - {206440 \over 81}
          + {4012 \over 3} ~\zeta_2
          - {968 \over 5} ~\zeta_2^2
          + 968 ~\zeta_3
          \Bigg)
\nonumber\\&&
       +  C_F^2 C_A^2   \Bigg(
          - {356573 \over 486}
          - 688 ~\zeta_2 ~\zeta_3
          + {129392 \over 27} ~\zeta_2
          - {15752 \over 15} ~\zeta_2^2
          + {57296 \over 9} ~\zeta_3
\nonumber\\&&
          - 1152 ~\zeta_5
          \Bigg)
       +  C_F^3 C_A   \Bigg(
           {34849 \over 18}
          - 2592 ~\zeta_2 ~\zeta_3
          - {22708 \over 9} ~\zeta_2
          + {2816 \over 15} ~\zeta_2^2
\nonumber\\&&
          + {14696 \over 3} ~\zeta_3
          \Bigg)
       +  C_F^4   \Bigg(
          - 3840 ~\zeta_2 ~\zeta_3
          - 5120 ~\zeta_3
          + 5376 ~\zeta_5
          \Bigg)
\Bigg]
\nonumber\\&&
+ {\cal \overline D}_3 \delta(1-z_1) \Bigg[
          n_f C_F C_A^2   \Bigg(
          - {7324 \over 27}
          + {352 \over 9} ~\zeta_2
          \Bigg)
       +  n_f C_F^2 C_A   \Bigg(
          - {133988 \over 243}
\nonumber\\&&
          + {14144 \over 27} ~\zeta_2
          + {448 \over 9} ~\zeta_3
          \Bigg)
       +  n_f C_F^3   \Bigg(
           {2764 \over 9}
          + {704 \over 9} ~\zeta_2
          + {5696 \over 9} ~\zeta_3
          \Bigg)
\nonumber\\&&
       +  n_f^2 C_F C_A   \Bigg(
           {1144 \over 27}
          - {32 \over 9} ~\zeta_2
          \Bigg)
       +  n_f^2 C_F^2   \Bigg(
           {7768 \over 243}
          - {1024 \over 27} ~\zeta_2
          \Bigg)
       +  n_f^3 C_F   \Bigg(
          - {160 \over 81}
          \Bigg)
\nonumber\\&&
       +  C_F C_A^3   \Bigg(
           {43648 \over 81}
          - {968 \over 9} ~\zeta_2
          \Bigg)
       +  C_F^2 C_A^2   \Bigg(
           {481216 \over 243}
          - {50272 \over 27} ~\zeta_2
          + {864 \over 5} ~\zeta_2^2
\nonumber\\&&
          - {8800 \over 9} ~\zeta_3
          \Bigg)
       +  C_F^3 C_A   \Bigg(
          - {26362 \over 9}
          - {1376 \over 9} ~\zeta_2
          + {1504 \over 5} ~\zeta_2^2
          - {19808 \over 9} ~\zeta_3
          \Bigg)
\nonumber\\&&
       +  C_F^4   \Bigg(
           1022
          + 976 ~\zeta_2
          - {1664 \over 5} ~\zeta_2^2
          - 480 ~\zeta_3
          \Bigg)
\Bigg]
\nonumber\\&&
+ {\cal \overline D}_4 \delta(1-z_1) \Bigg[
           n_f C_F C_A^2   \Bigg(
           {242 \over 9}
          \Bigg)
       +  n_f C_F^2 C_A   \Bigg(
           {5480 \over 27}
          - {80 \over 3} ~\zeta_2
          \Bigg)
\nonumber\\&&
       +  n_f C_F^3   \Bigg(
          - {440 \over 27}
          - {1040 \over 9} ~\zeta_2
          \Bigg)
       +  n_f^2 C_F C_A   \Bigg(
          - {44 \over 9}
          \Bigg)
       +  n_f^2 C_F^2   \Bigg(
          - {400 \over 27}
          \Bigg)
\nonumber\\&&
       +  n_f^3 C_F   \Bigg(
           {8 \over 27}
          \Bigg)
       +  C_F C_A^3   \Bigg(
          - {1331 \over 27}
          \Bigg)
       +  C_F^2 C_A^2   \Bigg(
          - {16780 \over 27}
          + {440 \over 3} ~\zeta_2
          \Bigg)
\nonumber\\&&
       +  C_F^3 C_A   \Bigg(
           {2480 \over 27}
          + {5720 \over 9} ~\zeta_2
          + 280 ~\zeta_3
          \Bigg)
       +  C_F^4   \Bigg(
           {2240 \over 3} ~\zeta_3
          \Bigg)
\Bigg]
\nonumber\\&&
+ {\cal \overline D}_5 \delta(1-z_1) \Bigg[
         n_f C_F^2 C_A   \Bigg(
          - {704 \over 27}
          \Bigg)
       +  n_f C_F^3   \Bigg(
          - {80 \over 3}
          \Bigg)
       +  n_f^2 C_F^2   \Bigg(
           {64 \over 27}
          \Bigg)
\nonumber\\&&
       +  C_F^2 C_A^2   \Bigg(
           {1936 \over 27}
          \Bigg)
       +  C_F^3 C_A   \Bigg(
           {536 \over 3}
          - 48 ~\zeta_2
          \Bigg)
       +  C_F^4   \Bigg(
          - 128
          - 96 ~\zeta_2
          \Bigg)
\Bigg]
\nonumber\\&&
+ {\cal \overline D}_6 \delta(1-z_1) \Bigg[
         n_f C_F^3   \Bigg(
           {56 \over 9}
          \Bigg)
       +  C_F^3 C_A   \Bigg(
          - {308 \over 9}
          \Bigg)
\Bigg]
+ {\cal \overline D}_7 \delta(1-z_1) \Bigg[
         C_F^4   \Bigg(
           {16 \over 3}
          \Bigg)
\Bigg]
\nonumber\\&& + (z_1 \leftrightarrow z_2) \,,
\end{eqnarray}
\begin{eqnarray}
   \Delta^{{\rm sv},(1)}_{Y,g} &=&
\delta(1-z_1) \delta(1-z_2) \Bigg[
        C_A   \Bigg(
           6 ~\zeta_2
          \Bigg)
\Bigg]
+ {\cal D}_0 {\cal \overline D}_0 \Bigg[
         C_A   \Bigg(
           2
          \Bigg)
\Bigg]
+ {\cal \overline D}_1 \delta(1-z_1)\Bigg[
          C_A   \Bigg(
           4
          \Bigg)
\Bigg]
\nonumber\\&& + (z_1 \leftrightarrow z_2) \,,
\nonumber\\[2ex]
   \Delta^{{\rm sv},(2)}_{Y,g} &=&
\delta(1-z_1) \delta(1-z_2)\Bigg[ 
           n_f C_F   \Bigg(
          - {67 \over 6}
          + 8 ~\zeta_3
          \Bigg)
       +  n_f C_A   \Bigg(
          - {40 \over 3}
          - {20 \over 3} ~\zeta_2
          - 4 ~\zeta_3
          \Bigg)
\nonumber\\&&
       +  C_A^2   \Bigg(
           {93 \over 2}
          + {134 \over 3} ~\zeta_2
          + {126 \over 5} ~\zeta_2^2
          - 22 ~\zeta_3
          \Bigg)
\Bigg]
+ {\cal D}_0 {\cal \overline D}_0 \Bigg[
         n_f C_A   \Bigg(
          - {20 \over 9}
          \Bigg)
       + C_A^2   \Bigg(
           {134 \over 9}
\nonumber\\&&
          + 4 ~\zeta_2
          \Bigg)
\Bigg]
+ {\cal D}_0 {\cal \overline D}_1 \Bigg[
        n_f C_A   \Bigg(
           {8 \over 3}
          \Bigg)
       +  C_A^2   \Bigg(
          - {44 \over 3}
          \Bigg)
\Bigg]
+ {\cal D}_0 {\cal \overline D}_2 \Bigg[
         C_A^2   \Bigg(
           24
          \Bigg)
\Bigg]
\nonumber\\&&
+ {\cal D}_1 {\cal \overline D}_1 \Bigg[C_A^2   \Bigg(
           24
          \Bigg)
\Bigg]
+ {\cal \overline D}_0 \delta(1-z_1) \Bigg[ 
         n_f C_A   \Bigg(
           {112 \over 27}
          - {8 \over 3} ~\zeta_2
          \Bigg)
       +  C_A^2   \Bigg(
          - {808 \over 27}
\nonumber\\&&
          + {44 \over 3} ~\zeta_2
          + 60 ~\zeta_3
          \Bigg)
\Bigg]
+ {\cal \overline D}_1 \delta(1-z_1) \Bigg[
         n_f C_A   \Bigg(
          - {40 \over 9}
          \Bigg)
       +  C_A^2   \Bigg(
           {268 \over 9}
          + 8 ~\zeta_2
          \Bigg)
\Bigg]
\nonumber\\&&
+ {\cal \overline D}_2 \delta(1-z_1) \Bigg[
         n_f C_A   \Bigg(
           {4 \over 3}
          \Bigg)
       +  C_A^2   \Bigg(
          - {22 \over 3}
          \Bigg)
\Bigg]
+ {\cal \overline D}_3 \delta(1-z_1)\Bigg[
         C_A^2   \Bigg(
           8
          \Bigg)
\Bigg]
\nonumber\\&& + (z_1 \leftrightarrow z_2)\,,
\nonumber\\[2ex]
   \Delta^{{\rm sv},(3)}_{Y,g} &=&
 {\cal D}_0 {\cal \overline D}_0 \Bigg[
         n_f C_F C_A   \Bigg(
          - 63
          + 48 ~\zeta_3
          \Bigg)
       +  n_f C_A^2   \Bigg(
          - {8422 \over 81}
          + {32 \over 3} ~\zeta_2
          + 16 ~\zeta_3
          \Bigg)
\nonumber\\&&
       +  n_f^2 C_A   \Bigg(
           {200 \over 81}
          - {16 \over 9} ~\zeta_2
          \Bigg)
       +  C_A^3   \Bigg(
           {30569 \over 81}
          + {52 \over 9} ~\zeta_2
          - {32 \over 5} ~\zeta_2^2
          - 352 ~\zeta_3
          \Bigg)
\Bigg]
\nonumber\\&&
+ {\cal D}_0 {\cal \overline D}_1 \Bigg[
         n_f C_F C_A   \Bigg(
           8
          \Bigg)
       +  n_f C_A^2   \Bigg(
           {3656 \over 27}
          - {224 \over 3} ~\zeta_2
          \Bigg)
       +  n_f^2 C_A   \Bigg(
          - {160 \over 27}
          \Bigg)
\nonumber\\&&
       +  C_A^3   \Bigg(
          - {16816 \over 27}
          + {1232 \over 3} ~\zeta_2
          + 976 ~\zeta_3
          \Bigg)
\Bigg]
+ {\cal D}_0 {\cal \overline D}_2 \Bigg[
         n_f C_A^2   \Bigg(
          - {656 \over 9}
          \Bigg)
\nonumber\\&&
       +  n_f^2 C_A   \Bigg(
           {16 \over 9}
          \Bigg)
       +  C_A^3   \Bigg(
           {3700 \over 9}
          - 192 ~\zeta_2
          \Bigg)
\Bigg]
+ {\cal D}_0 {\cal \overline D}_3 \Bigg[
          n_f C_A^2   \Bigg(
           {160 \over 9}
          \Bigg)
\nonumber\\&&
       +  C_A^3   \Bigg(
          - {880 \over 9}
          \Bigg)
\Bigg]
+ {\cal D}_0 {\cal \overline D}_4 \Bigg[C_A^3   \Bigg(
           40
          \Bigg)
\Bigg]
+ {\cal D}_1 {\cal \overline D}_1 \Bigg[
         n_f C_A^2   \Bigg(
          - {656 \over 9}
          \Bigg)
\nonumber\\&&
       +  n_f^2 C_A   \Bigg(
           {16 \over 9}
          \Bigg)
       +  C_A^3   \Bigg(
           {3700 \over 9}
          - 192 ~\zeta_2
          \Bigg)
\Bigg]
+ {\cal D}_1 {\cal \overline D}_2 \Bigg[
         n_f C_A^2   \Bigg(
           {160 \over 3}
          \Bigg)
\nonumber\\&&
       +  C_A^3   \Bigg(
          - {880 \over 3}
          \Bigg)
\Bigg]
+ {\cal D}_1 {\cal \overline D}_3 \Bigg[
         C_A^3   \Bigg(
           160
          \Bigg)
\Bigg]
+ {\cal D}_2 {\cal \overline D}_2 \Bigg[
         C_A^3   \Bigg(
           120
          \Bigg)
\Bigg]
\nonumber\\&&
+ {\cal \overline D}_0 \delta(1-z_1) \Bigg[ 
            n_f C_F C_A   \Bigg(
           {1711 \over 27}
          - 8 ~\zeta_2
          - {32 \over 5} ~\zeta_2^2
          - {304 \over 9} ~\zeta_3
          \Bigg)
\nonumber\\&&
       +  n_f C_A^2   \Bigg(
           {62626 \over 729}
          - {6416 \over 81} ~\zeta_2
          + {16 \over 5} ~\zeta_2^2
          - {392 \over 3} ~\zeta_3
          \Bigg)
       +  n_f^2 C_A   \Bigg(
          - {1856 \over 729}
\nonumber\\&&
          + {160 \over 27} ~\zeta_2
          - {32 \over 27} ~\zeta_3
          \Bigg)
       +  C_A^3   \Bigg(
          - {297029 \over 729}
          - {608 \over 3} ~\zeta_2 ~\zeta_3
          + {18056 \over 81} ~\zeta_2
\nonumber\\&&
          + {88 \over 5} ~\zeta_2^2
          + {27128 \over 27} ~\zeta_3
          + 192 ~\zeta_5
          \Bigg)
\Bigg]
+ {\cal \overline D}_1 \delta(1-z_1)\Bigg[
          n_f C_F C_A   \Bigg(
          - 126
          + 96 ~\zeta_3
          \Bigg)
\nonumber\\&&
       +  n_f C_A^2   \Bigg(
          - {16844 \over 81}
          + {64 \over 3} ~\zeta_2
          + 32 ~\zeta_3
          \Bigg)
       +  n_f^2 C_A   \Bigg(
           {400 \over 81}
          - {32 \over 9} ~\zeta_2
          \Bigg)
\nonumber\\&&
       +  C_A^3   \Bigg(
           {61138 \over 81}
          + {104 \over 9} ~\zeta_2
          - {64 \over 5} ~\zeta_2^2
          - 704 ~\zeta_3
          \Bigg)
\Bigg]
+ {\cal \overline D}_2 \delta(1-z_1) \Bigg[
         n_f C_F C_A   \Bigg(
           4
          \Bigg)
\nonumber\\&&
       +  n_f C_A^2   \Bigg(
           {1828 \over 27}
          - {112 \over 3} ~\zeta_2
          \Bigg)
       +  n_f^2 C_A   \Bigg(
          - {80 \over 27}
          \Bigg)
       +  C_A^3   \Bigg(
          - {8408 \over 27}
          + {616 \over 3} ~\zeta_2
\nonumber\\&&
          + 488 ~\zeta_3
          \Bigg)
\Bigg]
+ {\cal \overline D}_3 \delta(1-z_1) \Bigg[
         n_f C_A^2   \Bigg(
          - {656 \over 27}
          \Bigg)
       +  n_f^2 C_A   \Bigg(
           {16 \over 27}
          \Bigg)
       +  C_A^3   \Bigg(
           {3700 \over 27}
\nonumber\\&&
          - 64 ~\zeta_2
          \Bigg)
\Bigg]
+ {\cal \overline D}_4 \delta(1-z_1)\Bigg[
         n_f C_A^2   \Bigg(
           {40 \over 9}
          \Bigg)
       +  C_A^3   \Bigg(
          - {220 \over 9}
          \Bigg)
\Bigg]
\nonumber\\&&
+ {\cal \overline D}_5 \delta(1-z_1)\Bigg[ 
          C_A^3   \Bigg(
           8
          \Bigg)
\Bigg]
\nonumber\\&& + (z_1 \leftrightarrow z_2) \,,
\nonumber\\[2ex]
   \Delta^{{\rm sv},(4)}_{Y,g} &=&
{\cal D}_0 {\cal \overline D}_1 \Bigg[
           n_f C_F C_A^2   \Bigg(
           1656
          - 224 ~\zeta_2
          - {384 \over 5} ~\zeta_2^2
          - 992 ~\zeta_3
          \Bigg)
       +  n_f C_F^2 C_A   \Bigg(
          - 4
          \Bigg)
\nonumber\\&&
       +  n_f C_A^3   \Bigg(
           {1147774 \over 243}
          - {120848 \over 27} ~\zeta_2
          + {2304 \over 5} ~\zeta_2^2
          - {47648 \over 9} ~\zeta_3
          \Bigg)
\nonumber\\&&
       +  n_f^2 C_F C_A   \Bigg(
          - {1400 \over 9}
          + {320 \over 3} ~\zeta_3
          \Bigg)
       +  n_f^2 C_A^2   \Bigg(
          - {109190 \over 243}
          + {3296 \over 9} ~\zeta_2
          + {1088 \over 9} ~\zeta_3
          \Bigg)
\nonumber\\&&
       +  n_f^3 C_A   \Bigg(
           {800 \over 81}
          - {64 \over 9} ~\zeta_2
          \Bigg)
       +  C_A^4   \Bigg(
          - {3407840 \over 243}
          - 14240 ~\zeta_2 ~\zeta_3
          + {397592 \over 27} ~\zeta_2
\nonumber\\&&
          - 2112 ~\zeta_2^2
          + 30448 ~\zeta_3
          + 8448 ~\zeta_5
          \Bigg)
\Bigg]
+ {\cal D}_0 {\cal \overline D}_2 \Bigg[
           n_f C_F C_A^2   \Bigg(
          - {3148 \over 3}
          + 768 ~\zeta_3
          \Bigg)
\nonumber\\&&
       +  n_f C_A^3   \Bigg(
          - {271148 \over 81}
          + {18080 \over 9} ~\zeta_2
          + 1408 ~\zeta_3
          \Bigg)
       +  n_f^2 C_F C_A   \Bigg(
           {40 \over 3}
          \Bigg)
\nonumber\\&&
       +  n_f^2 C_A^2   \Bigg(
           {19288 \over 81}
          - {1120 \over 9} ~\zeta_2
          \Bigg)
       +  n_f^3 C_A   \Bigg(
          - {160 \over 27}
          \Bigg)
       +  C_A^4   \Bigg(
           {862648 \over 81}
          - {72472 \over 9} ~\zeta_2
\nonumber\\&&
          + {2112 \over 5} ~\zeta_2^2
          - 11968 ~\zeta_3
          \Bigg)
\Bigg]
+ {\cal D}_0 {\cal \overline D}_3 \Bigg[
          n_f C_F C_A^2   \Bigg(
           {160 \over 3}
          \Bigg)
       +  n_f C_A^3   \Bigg(
           {3256 \over 3}
\nonumber\\&&
          - {5120 \over 9} ~\zeta_2
          \Bigg)
       +  n_f^2 C_A^2   \Bigg(
          - {2128 \over 27}
          \Bigg)
       +  n_f^3 C_A   \Bigg(
           {32 \over 27}
          \Bigg)
       +  C_A^4   \Bigg(
          - {104764 \over 27}
\nonumber\\&&
          + {28160 \over 9} ~\zeta_2
          + {12320 \over 3} ~\zeta_3
          \Bigg)
\Bigg]
+ {\cal D}_0 {\cal \overline D}_4 \Bigg[
          n_f C_A^3   \Bigg(
          - {7120 \over 27}
          \Bigg)
       +  n_f^2 C_A^2   \Bigg(
           {320 \over 27}
          \Bigg)
\nonumber\\&&
       +  C_A^4   \Bigg(
           {33800 \over 27}
          - 720 ~\zeta_2
          \Bigg)
\Bigg]
+ {\cal D}_0 {\cal \overline D}_5 \Bigg[
          n_f C_A^3   \Bigg(
           {112 \over 3}
          \Bigg)
       +  C_A^4   \Bigg(
          - {616 \over 3}
          \Bigg)
\Bigg]
\nonumber\\&&
+ {\cal D}_0 {\cal \overline D}_6 \Bigg[
          C_A^4   \Bigg(
           {112 \over 3}
          \Bigg)
\Bigg]
+ {\cal D}_1 {\cal \overline D}_1 \Bigg[
          n_f C_F C_A^2   \Bigg(
          - {3148 \over 3}
          + 768 ~\zeta_3
          \Bigg)
\nonumber\\&&
       +  n_f C_A^3   \Bigg(
          - {271148 \over 81}
          + {18080 \over 9} ~\zeta_2
          + 1408 ~\zeta_3
          \Bigg)
       +  n_f^2 C_F C_A   \Bigg(
           {40 \over 3}
          \Bigg)
\nonumber\\&&
       +  n_f^2 C_A^2   \Bigg(
           {19288 \over 81}
          - {1120 \over 9} ~\zeta_2
          \Bigg)
       +  n_f^3 C_A   \Bigg(
          - {160 \over 27}
          \Bigg)
       +  C_A^4   \Bigg(
           {862648 \over 81}
          - {72472 \over 9} ~\zeta_2
\nonumber\\&&
          + {2112 \over 5} ~\zeta_2^2
          - 11968 ~\zeta_3
          \Bigg)
\Bigg]
+ {\cal D}_1 {\cal \overline D}_2 \Bigg[
          n_f C_F C_A^2   \Bigg(
           160
          \Bigg)
       +  n_f C_A^3   \Bigg(
           3256
\nonumber\\&&
          - {5120 \over 3} ~\zeta_2
          \Bigg)
       +  n_f^2 C_A^2   \Bigg(
          - {2128 \over 9}
          \Bigg)
       +  n_f^3 C_A   \Bigg(
           {32 \over 9}
          \Bigg)
       +  C_A^4   \Bigg(
          - {104764 \over 9}
          + {28160 \over 3} ~\zeta_2
\nonumber\\&&
          + 12320 ~\zeta_3
          \Bigg)
\Bigg]
+ {\cal D}_1 {\cal \overline D}_3 \Bigg[
          n_f C_A^3   \Bigg(
          - {28480 \over 27}
          \Bigg)
       +  n_f^2 C_A^2   \Bigg(
           {1280 \over 27}
          \Bigg)
       +  C_A^4   \Bigg(
           {135200 \over 27}
\nonumber\\&&
          - 2880 ~\zeta_2
          \Bigg)
\Bigg]
+ {\cal D}_1 {\cal \overline D}_4 \Bigg[
         n_f C_A^3   \Bigg(
           {560 \over 3}
          \Bigg)
       +  C_A^4   \Bigg(
          - {3080 \over 3}
          \Bigg)
\Bigg]
+ {\cal D}_1 {\cal \overline D}_5 \Bigg[
          C_A^4   \Bigg(
           224
          \Bigg)
\Bigg]
\nonumber\\&&
+ {\cal D}_2 {\cal \overline D}_2 \Bigg[
          n_f C_A^3   \Bigg(
          - {7120 \over 9}
          \Bigg)
       +  n_f^2 C_A^2   \Bigg(
           {320 \over 9}
          \Bigg)
       +  C_A^4   \Bigg(
           {33800 \over 9}
          - 2160 ~\zeta_2
          \Bigg)
\Bigg]
\nonumber\\&&
+ {\cal D}_2 {\cal \overline D}_3 \Bigg[
          n_f C_A^3   \Bigg(
           {1120 \over 3}
          \Bigg)
       +  C_A^4   \Bigg(
          - {6160 \over 3}
          \Bigg)
\Bigg]
+ {\cal D}_2 {\cal \overline D}_4 \Bigg[
         C_A^4   \Bigg(
           560
          \Bigg)
\Bigg]
\nonumber\\&&
+ {\cal D}_3 {\cal \overline D}_3 \Bigg[
         C_A^4   \Bigg(
           {1120 \over 3}
          \Bigg)
\Bigg]
+ {\cal \overline D}_2 \delta(1-z_1) \Bigg[ 
            n_f C_F C_A^2   \Bigg(
           828
          - 112 ~\zeta_2
          - {192 \over 5} ~\zeta_2^2
\nonumber\\&&
          - 496 ~\zeta_3
          \Bigg)
       +  n_f C_F^2 C_A   \Bigg(
          - 2
          \Bigg)
       +  n_f C_A^3   \Bigg(
           {573887 \over 243}
          - {60424 \over 27} ~\zeta_2
          + {1152 \over 5} ~\zeta_2^2
\nonumber\\&&
          - {23824 \over 9} ~\zeta_3
          \Bigg)
       +  n_f^2 C_F C_A   \Bigg(
          - {700 \over 9}
          + {160 \over 3} ~\zeta_3
          \Bigg)
       +  n_f^2 C_A^2   \Bigg(
          - {54595 \over 243}
\nonumber\\&&
          + {1648 \over 9} ~\zeta_2
          + {544 \over 9} ~\zeta_3
          \Bigg)
       +  n_f^3 C_A   \Bigg(
           {400 \over 81}
          - {32 \over 9} ~\zeta_2
          \Bigg)
       +  C_A^4   \Bigg(
          - {1703920 \over 243}
\nonumber\\&&
          - 7120 ~\zeta_2 ~\zeta_3
          + {198796 \over 27} ~\zeta_2
          - 1056 ~\zeta_2^2
          + 15224 ~\zeta_3
          + 4224 ~\zeta_5
          \Bigg)
\Bigg]
\nonumber\\&&
+ {\cal \overline D}_3 \delta(1-z_1) \Bigg[
            n_f C_F C_A^2   \Bigg(
          - {3148 \over 9}
          + 256 ~\zeta_3
          \Bigg)
       +  n_f C_A^3   \Bigg(
          - {271148 \over 243}
\nonumber\\&&
          + {18080 \over 27} ~\zeta_2
          + {1408 \over 3} ~\zeta_3
          \Bigg)
       +  n_f^2 C_F C_A   \Bigg(
           {40 \over 9}
          \Bigg)
       +  n_f^2 C_A^2   \Bigg(
           {19288 \over 243}
\nonumber\\&&
          - {1120 \over 27} ~\zeta_2
          \Bigg)
       +  n_f^3 C_A   \Bigg(
          - {160 \over 81}
          \Bigg)
       +  C_A^4   \Bigg(
           {862648 \over 243}
          - {72472 \over 27} ~\zeta_2
          + {704 \over 5} ~\zeta_2^2
\nonumber\\&&
          - {11968 \over 3} ~\zeta_3
          \Bigg)
\Bigg]
+ {\cal \overline D}_4 \delta(1-z_1) \Bigg[
          n_f C_F C_A^2   \Bigg(
           {40 \over 3}
          \Bigg)
       +  n_f C_A^3   \Bigg(
           {814 \over 3}
          - {1280 \over 9} ~\zeta_2
          \Bigg)
\nonumber\\&&
       +  n_f^2 C_A^2   \Bigg(
          - {532 \over 27}
          \Bigg)
       +  n_f^3 C_A   \Bigg(
           {8 \over 27}
          \Bigg)
       +  C_A^4   \Bigg(
          - {26191 \over 27}
          + {7040 \over 9} ~\zeta_2
          + {3080 \over 3} ~\zeta_3
          \Bigg)
\Bigg]
\nonumber\\&&
+ {\cal \overline D}_5 \delta(1-z_1) \Bigg[
         n_f C_A^3   \Bigg(
          - {1424 \over 27}
          \Bigg)
       +  n_f^2 C_A^2   \Bigg(
           {64 \over 27}
          \Bigg)
       +  C_A^4   \Bigg(
           {6760 \over 27}
          - 144 ~\zeta_2
          \Bigg)
\Bigg]
\nonumber\\&&
+ {\cal \overline D}_6 \delta(1-z_1) \Bigg[
          n_f C_A^3   \Bigg(
           {56 \over 9}
          \Bigg)
       +  C_A^4   \Bigg(
          - {308 \over 9}
          \Bigg)
\Bigg]
+ {\cal \overline D}_7 \delta(1-z_1) \Bigg[
         C_A^4   \Bigg(
           {16 \over 3}
          \Bigg)
\Bigg]
\nonumber\\&& + (z_1 \leftrightarrow z_2) \,,
\end{eqnarray}
%
%
%
where
\begin{eqnarray}
{\cal D}_i=\left[{\log^i(1-z_1) \over (1-z_1)}\right]_+~,
\quad \quad \quad \quad {\cal \overline D}_i=
\left[{\log^i(1-z_2) \over (1-z_2)}\right]_+ \,.
\end{eqnarray}



\begin{thebibliography}{99}
\bibitem{Dittmar:2005ed}
  M.~Dittmar {\it et al.},
  [arXiv:hep-ph/0511119].

\bibitem{Drell:1970wh}
S.~D.~Drell and T.~M.~Yan,
Phys.\ Rev.\ Lett.\  {\bf 25} (1970) 316
[Erratum-ibid.\  {\bf 25}, 902 (1970)].

\bibitem{Affolder:2000rx}
A.~A.~Affolder {\it et al.}  [CDF Collaboration],
  Phys.\ Rev.\ D {\bf 63} (2001) 011101
  [arXiv:hep-ex/0006025].

\bibitem{Abe:1998rv}
  F.~Abe {\it et al.}  [CDF Collaboration],
  Phys.\ Rev.\ Lett.\  {\bf 81} (1988) 5754
  [arXiv:hep-ex/9809001].

\bibitem{Affolder:2001ha}
A.~A.~Affolder {\it et al.}  [CDF Collaboration],
     Phys.\ Rev.\ Lett.\  {\bf 87} (2001) 131802
       [arXiv:hep-ex/0106047].

\bibitem{Patwa:2006rd}
  A.~Patwa  [CDF Collaboration],
    [arXiv:hep-ex/0605082].

\bibitem{Djouadi:2005gi}
  A.~Djouadi,
  %
  [arXiv:hep-ph/0503172].

\bibitem{Djouadi:2005gj}
  A.~Djouadi,
  %
  [arXiv:hep-ph/0503173].

\bibitem{Barate:2003sz}
R.~Barate {\it et al.}  [LEP Working Group for Higgs boson searches],
  Phys.\ Lett.\ B {\bf 565} (2003) 61
  [arXiv:hep-ex/0306033].

\bibitem{:2004qh}
   [LEP Collaborations],
  [arXiv:hep-ex/0412015],
The LEP Collaborations ALEPH, DeLPHI, L3, OPAL, 
the LEP Electroweak Working Group, the SLD Electroweak and Heavy Flavor Groups. 
(URL: http://lepewwg.web.cern.ch/LEPEWWG/)

\bibitem{Kubar-Andre:1978uy}
  J.~Kubar-Andre and F.~E.~Paige,
      Phys.\ Rev.\ D {\bf 19} (1979) 221.
\bibitem{Altarelli:1978id}
  G.~Altarelli, R.~K.~Ellis and G.~Martinelli,
  Nucl.\ Phys.\ B {\bf 143} (1978) 521,
  [Erratum-ibid.\ B {\bf 146} (1978) 544].

\bibitem{Humpert:1980uv}
B.~Humpert and W.~L.~van Neerven,
Nucl.\ Phys.\ B {\bf 184} (1981) 225.

\bibitem{Dawson:1990zj}
  S.~Dawson,
  %
  Nucl.\ Phys.\ B {\bf 359} (1991) 283.
                                                                                
\bibitem{Djouadi:1991tk}
  A.~Djouadi, M.~Spira and P.~M.~Zerwas,
  %
  Phys.\ Lett.\ B {\bf 264} (1991) 440.
                                                                                
\bibitem{Spira:1995rr}
  M.~Spira, A.~Djouadi, D.~Graudenz and P.~M.~Zerwas,
  %
  Nucl.\ Phys.\ B {\bf 453} (1995) 17
  [arXiv:hep-ph/9504378].

\bibitem{Matsuura:1987wt}
  T.~Matsuura and W.~L.~van Neerven,
  %
  Z.\ Phys.\ C {\bf 38} (1988) 623.
                                                                                
\bibitem{Matsuura:1988sm}
  T.~Matsuura, S.~C.~van der Marck and W.~L.~van Neerven,
  %
  Nucl.\ Phys.\ B {\bf 319}, 570 (1989).

\bibitem{Hamberg:1990np}
  R.~Hamberg, W.~L.~van Neerven and T.~Matsuura,
  %
  Nucl.\ Phys.\ B {\bf 359} (1991) 343,
  [Erratum-ibid.\ B {\bf 644} (2002) 403].

\bibitem{Harlander:2001is}
  R.~V.~Harlander and W.~B.~Kilgore,
  %
  Phys.\ Rev.\ D {\bf 64} (2001) 013015
  [arXiv:hep-ph/0102241].

\bibitem{Catani:2001ic}
  S.~Catani, D.~de Florian and M.~Grazzini,
  %
  JHEP {\bf 0105} (2001) 025
  [arXiv:hep-ph/0102227].

\bibitem{Catani:2003zt}
  S.~Catani, D.~de Florian, M.~Grazzini and P.~Nason,
  JHEP {\bf 0307} (2003) 028
  [arXiv:hep-ph/0306211].

\bibitem{Harlander:2002wh}
  R.~V.~Harlander and W.~B.~Kilgore,
  %
  Phys.\ Rev.\ Lett.\  {\bf 88} (2002) 201801
  [arXiv:hep-ph/0201206].

\bibitem{Anastasiou:2002yz}
  C.~Anastasiou and K.~Melnikov,
  %
  Nucl.\ Phys.\ B {\bf 646} (2002) 220
  [arXiv:hep-ph/0207004].

\bibitem{Ravindran:2003um}
  V.~Ravindran, J.~Smith and W.~L.~van Neerven,
  %
  Nucl.\ Phys.\ B {\bf 665} (2003) 325
  [arXiv:hep-ph/0302135].

\bibitem{Harlander:2003ai}
  R.~V.~Harlander and W.~B.~Kilgore,
  Phys.\ Rev.\ D {\bf 68} (2003) 013001
  [arXiv:hep-ph/0304035].

\bibitem{Ravindran:2004mb}
  V.~Ravindran, J.~Smith and W.~L.~van Neerven,
  %
  Nucl.\ Phys.\ B {\bf 704} (2005) 332
  [arXiv:hep-ph/0408315].

\bibitem{Sterman:1986aj}
  G.~Sterman,
  %
  Nucl.\ Phys.\ B {\bf 281} (1987) 310.

\bibitem{Catani:1989ne}
  S.~Catani and L.~Trentadue,
  %
  Nucl.\ Phys.\ B {\bf 327} (1989) 323.

\bibitem{Kodaira:1981nh}
  J.~Kodaira and L.~Trentadue,
  Phys.\ Lett.\ B {\bf 112} (1982) 66.

\bibitem{Vogt:2000ci}
  A.~Vogt,
  %
  Phys.\ Lett.\ B {\bf 497} (2001) 228
  [arXiv:hep-ph/0010146].

\bibitem{Moch:2004pa}
  S.~Moch, J.~A.~M.~Vermaseren and A.~Vogt,
  %
  Nucl.\ Phys.\ B {\bf 688} (2004) 101
  [arXiv:hep-ph/0403192].

\bibitem{Vogt:2004mw}
  A.~Vogt, S.~Moch and J.~A.~M.~Vermaseren,
  %
  Nucl.\ Phys.\ B {\bf 691} (2004) 129
  [arXiv:hep-ph/0404111].

\bibitem{Moch:2005id}
  S.~Moch, J.~A.~M.~Vermaseren and A.~Vogt,
  %
  JHEP {\bf 0508} (2005) 049
  [arXiv:hep-ph/0507039].

\bibitem{Moch:2005tm}
  S.~Moch, J.~A.~M.~Vermaseren and A.~Vogt,
  %
  Phys.\ Lett.\ B {\bf 625} (2005) 245
  [arXiv:hep-ph/0508055].


\bibitem{Vermaseren:2005qc}
  J.~A.~M.~Vermaseren, A.~Vogt and S.~Moch,
  Nucl.\ Phys.\ B {\bf 724}, 3 (2005)
  [arXiv:hep-ph/0504242].

\bibitem{Moch:2005ba}
  S.~Moch, J.~A.~M.~Vermaseren and A.~Vogt,
  %
  Nucl.\ Phys.\ B {\bf 726} (2005) 317
  [arXiv:hep-ph/0506288].

\bibitem{Blumlein:2004xt}
  J.~Blumlein and J.~A.~M.~Vermaseren,
  Phys.\ Lett.\ B {\bf 606} (2005) 130
  [arXiv:hep-ph/0411111].

\bibitem{Moch:2005ky}
  S.~Moch and A.~Vogt,
  %
  Phys.\ Lett.\ B {\bf 631} (2005) 48
  [arXiv:hep-ph/0508265].

\bibitem{Laenen:2005uz}
  E.~Laenen and L.~Magnea,
  Phys.\ Lett.\ B {\bf 632} (2006) 270 
  [arXiv:hep-ph/0508284].

\bibitem{Idilbi:2005ni}
  A.~Idilbi, X.~d.~Ji, J.~P.~Ma and F.~Yuan,
   Phys.\ Rev.\ D {\bf 73} (2006)  077501
  [arXiv:hep-ph/0509294].

\bibitem{Ravindran:2005vv}
  V.~Ravindran,
  Nucl.\ Phys.\ B {\bf 746} (2006) 58
  [arXiv:hep-ph/0512249].

\bibitem{Ravindran:2006cg}
  V.~Ravindran,
  Nucl.\ Phys.\ B {\bf 752} (2006) 173
      [arXiv:hep-ph/0603041].

\bibitem{Blumlein:2006pj}
J.~Blumlein and V.~Ravindran,
   [arXiv:hep-ph/0605011].


\bibitem{Blumlein:2000wh}
  J.~Blumlein, V.~Ravindran and W.~L.~van Neerven,
  Nucl.\ Phys.\ B {\bf 586} (2000) 349
  [arXiv:hep-ph/0004172].

\bibitem{Blumlein:2005im}
  J.~Blumlein and V.~Ravindran,
  Nucl.\ Phys.\ B {\bf 716} (2005) 128
  [arXiv:hep-ph/0501178].

\bibitem{Dokshitzer:2005bf}
  Y.~L.~Dokshitzer, G.~Marchesini and G.~P.~Salam,
  [arXiv:hep-ph/0511302].

\bibitem{Cafarella:2005nb}
  A.~Cafarella, C.~Coriano, M.~Guzzi and J.~Smith,
  arXiv:hep-ph/0510179.

\bibitem{laenensterman}
E. Laenen and G. Sterman, FERMILAB-CONF-92/359-T,
Contribution to the Fermilab Meeting, DPF 1992 World Scientific
Vol 2 edited by C.H. Albright, P.H. Kasper, R. Raja and J. Yoh, (1993).

\bibitem{Rijken:1994sh}
P.~J.~Rijken and W.~L.~van Neerven,
   Phys.\ Rev.\ D {\bf 51} (1995) 44
  [arXiv:hep-ph/9408366].


\bibitem{Mathews:2004xp}
P.~Mathews, V.~Ravindran, K.~Sridhar and W.~L.~van Neerven,
   Nucl.\ Phys.\ B {\bf 713} (2005) 333
  [arXiv:hep-ph/0411018].

\bibitem{Chetyrkin:1997un}
  K.~G.~Chetyrkin, B.~A.~Kniehl and M.~Steinhauser,
  %
  Nucl.\ Phys.\ B {\bf 510} (1998) 61
  [arXiv:hep-ph/9708255].


\bibitem{vanRitbergen:1997va}
  T.~van Ritbergen, J.~A.~M.~Vermaseren and S.~A.~Larin,
  %
  Phys.\ Lett.\ B {\bf 400} (1997) 379
  [arXiv:hep-ph/9701390];

\bibitem{Sudakov:1954sw}
  V.~V.~Sudakov,
  %
  Sov.\ Phys.\ JETP {\bf 3} (1956) 65
  [Zh.\ Eksp.\ Teor.\ Fiz.\  {\bf 30} (1956) 87].

\bibitem{Mueller:1979ih}
  A.~H.~Mueller,
  %
  Phys.\ Rev.\ D {\bf 20} (1979) 2037.

\bibitem{Collins:1980ih}
  J.~C.~Collins,
  %
  Phys.\ Rev.\ D {\bf 22} (1980) 1478.

\bibitem{Sen:1981sd}
  A.~Sen,
  %
  Phys.\ Rev.\ D {\bf 24} (1981) 3281.

\bibitem{Aybat:2006mz}
 S.~M.~Aybat, L.~J.~Dixon and G.~Sterman,
 Phys.\ Rev.\ Lett.\ {\bf 99} (2006) 072001 
 [arXiv:hep-ph/0607309].

\bibitem{Aybat:2006wq}
 S.~M. Aybat, L.~J.~Dixon and G.~Sterman,
[arXiv:hep-ph/0606254].


\bibitem{Contopanagos:1996nh}
  H.~Contopanagos, E.~Laenen and G.~Sterman,
  %
  Nucl.\ Phys.\ B {\bf 484} (1997) 303
  [arXiv:hep-ph/9604313].

\bibitem{Eynck:2003fn}
  T.~O.~Eynck, E.~Laenen and L.~Magnea,
  %
  JHEP {\bf 0306} (2003) 057
  [arXiv:hep-ph/0305179].

\bibitem{ac}
L. Alvero and H. Contopanagos, Nucl. Phys. B {\bf 456} (1995) 497
[arXiv:hrp-ph/9411294].

\bibitem{Martin:2002dr}
  A.~D.~Martin, R.~G.~Roberts, W.~J.~Stirling and R.~S.~Thorne,
  %
  Phys.\ Lett.\ B {\bf 531} (2002) 216
  [arXiv:hep-ph/0201127].

\bibitem{Martin:2001es}
  A.~D.~Martin, R.~G.~Roberts, W.~J.~Stirling and R.~S.~Thorne,
  %
  Eur.\ Phys.\ J.\ C {\bf 23} (2002) 73
  [arXiv:hep-ph/0110215].

\bibitem{fgk}
D. de Florian, M. Grazzini, Z. Kunszt, Phys. Rev. Lett. {\bf 82} (1999)
5209, 
[arXiv:hep-ph/9902483].

\bibitem{rasm2}
V. Ravindran, J. Smith, W.L. van Neerven, Nucl. Phys. B {\bf 634} (2002) 247,
[arXiv:hep-ph/0201114].

\bibitem{glosser}
C.J. Glosser, C.J. Schmidt, JHEP {\bf 0212} (2002) 016, 
[arXiv:hep-ph/0209248].

\bibitem{fism}
B. Field, J. Smith, M.E. Tejeda-Yeomans, W.L. van Neerven, Phys. Lett.
B {\bf 551} (2003) 137, 
[arXiv:hep-ph/0210369].

\bibitem{Anastasiou:2002qz}
C.~Anastasiou, L.~J.~Dixon and K.~Melnikov,
Nucl.\ Phys.\ Proc.\ Suppl.\  {\bf 116} (2003) 193
[arXiv:hep-ph/0211141].

\bibitem{boca}
G. Bozzi, S. Catani, D. de Florian, M. Grazzini, Phys. Lett. B {\bf 564}
(2003) 65,
[arXiv:hep-ph/0302104].

\bibitem{Anastasiou:2003ds}
 C.~Anastasiou, L.~J.~Dixon, K.~Melnikov and F.~Petriello,
Phys.\ Rev.\ D {\bf 69} (2004) 094008
[arXiv:hep-ph/0312266].

\bibitem{Anastasiou:2003yy}
 C.~Anastasiou, L.~J.~Dixon, K.~Melnikov and F.~Petriello,
 Phys.\ Rev.\ Lett.\  {\bf 91} (2003) 182002
[arXiv:hep-ph/0306192].



\bibitem{field}
B. Field, Phys. Rev. D {\bf 70} (2004) 054008 
[arXiv:hep-ph/0405219].














\end{thebibliography}
\end{document}